\documentclass[twocolumn,tighten]{aastex63}
\accepted{}
\received{}
\submitjournal{AAS journals}
\usepackage{graphicx}
\usepackage{natbib}
\usepackage{latexsym}
\usepackage{amssymb}
\usepackage{longtable}
\usepackage{amsmath}
\usepackage{url}
\def\msun{\ifmmode {\rm\,M_\odot}\else ${\rm\,M_\odot}$\fi}
\def\Msun{\ifmmode {\rm\,\it{M_\odot}}\else ${\rm\,M_\odot}$\fi}
\def\lsun{\ifmmode {\rm\,L_\odot}\else ${\rm\,L_\odot}$\fi}
\def\Lsun{\ifmmode {\rm\,\it{L_\odot}}\else ${\rm\,L_\odot}$\fi}
\def\rsun{\ifmmode {\rm\,R_\odot}\else ${\rm\,R_\odot}$\fi}
\def\Rsun{\ifmmode {\rm\,\it{R_\odot}}\else ${\rm\,R_\odot}$\fi}
\def\Tsun{\ifmmode {\rm\,T_\odot}\else ${\rm\,T_\odot}$\fi}
\def\arcsec{\ifmmode {^{\prime\prime}}\else $^{\prime\prime}$\fi}
\def\asec{\ifmmode {^{\prime\prime}}\else $^{\prime\prime}$\fi}
\def\arcmin{\ifmmode {^{\prime}}\else $^{\prime}$\fi}
\def\amin{\ifmmode {^{\prime}}\else $^{\prime}$\fi}
\def\simlt{\mathrel{\spose{\lower 3pt\hbox{$\mathchar"218$}}
     \raise 2.0pt\hbox{$\mathchar"13C$}}}
\def\simgt{\mathrel{\spose{\lower 3pt\hbox{$\mathchar"218$}}
\     \raise 2.0pt\hbox{$\mathchar"13E$}}}

\def\escape~{\textit{ESCAPE}}

\begin{document}

\author[0000-0002-1002-3674]{Kevin France}
\affiliation{Laboratory for Atmospheric and Space Physics, University of Colorado Boulder, Boulder, CO 80303}

\author[0000-0003-2631-5265]{Nicole Arulanantham}
\affiliation{Space Telescope Science Institute, 3700 San Martin Drive, Baltimore, MD 21218, USA}

\author{Erin Maloney}
\affiliation{Laboratory for Atmospheric and Space Physics, University of Colorado Boulder, Boulder, CO 80303}

\author{P. Wilson Cauley}
\affiliation{Laboratory for Atmospheric and Space Physics, University of Colorado Boulder, Boulder, CO 80303}

\author[0000-0001-6015-646X]{P. \'Abrah\'am} \affiliation{Konkoly Observatory, Research Centre for Astronomy and Earth Sciences, E\"otv\"os Lor\'and Research Network (ELKH), Konkoly-Thege Mikl\'os \'ut 15-17, 1121 Budapest, Hungary} \affiliation{CSFK, MTA Centre of Excellence, Budapest, Konkoly Thege Miklós út 15-17., H-1121, Hungary} \affiliation{ELTE Eötvös Loránd University, Institute of Physics, Pázmány Péter sétány 1/A, H-1117 Budapest, Hungary}

\author{Juan M. Alcal\'a}
\affiliation{INAF-Osservatorio Astronomico di Capodimonte, via Moiariello 16, 80131 Napoli, Italy}

\author[0000-0002-3913-3746]{Justyn Campbell-White}
\affiliation{European Southern Observatory, Karl-Schwarzschild-Strasse 2, 85748 Garching bei M\"unchen, Germany}

\author[0000-0002-5261-6216]{Eleonora Fiorellino}
\affiliation{INAF-Osservatorio Astronomico di Capodimonte, via Moiariello 16, 80131 Napoli, Italy}

\author[0000-0002-7154-6065]{Gregory J. Herczeg}
\affiliation{Kavli Institute for Astronomy and Astrophysics, Peking University, Beijing 100871, China}
\affiliation{Department of Astronomy, Peking University, Beijing 100871, China}

\author[0000-0002-9190-0113]{Brunella Nisini}
\affiliation{ INAF - Osservtorio Astronomico di Roma. Via di Frascati 33, 00078 Monte Porzio Catone, Italy}

\author[0000-0002-4147-3846]{Miguel Vioque}
\affiliation{Joint ALMA Observatory, Alonso de Córdova 3107, Vitacura, Santiago 763-0355, Chile}
\affiliation{National Radio Astronomy Observatory, 520 Edgemont Road, Charlottesville, VA 22903, USA}

\correspondingauthor{Kevin France}
\email{kevin.france@colorado.edu}

\title{The Radial Distribution and Excitation of H$_{2}$ around Young Stars in the $HST$-ULLYSES Survey}

\begin{abstract} 

The spatial distribution and evolution of gas in the inner 10 au of protoplanetary disks form the basis for estimating the initial conditions of planet formation. Among the most important constraints derived from spectroscopic observations of the inner disk are the radial distributions of the major gas phase constituents, how the properties of the gas change with inner disk dust evolution, and how chemical abundances and excitation conditions are influenced by the high-energy radiation from the central star.  We present a survey of the radial distribution, excitation, and evolution of inner disk molecular hydrogen (H$_{2}$) obtained as part of the $HST$/ULLYSES program.  We analyze far-ultraviolet spectroscopy of 71 (63 accreting) pre-main sequence systems in the ULLYSES DR5 release to characterize the H$_{2}$ emission lines, H$_{2}$ dissociation continuum emission, and major photochemical/disk evolution driving UV emissions (Ly$\alpha$, UV continuum, and C IV).  We use the widths of the H$_{2}$ emission lines to show that most fluorescent H$_{2}$ arises between 0.1 – 1.4 au from the parent star, and show positive correlations of the average emitting radius with the accretion luminosity and with the dust disk mass.  We find a strong correlation between H$_{2}$ dissociation emission and both the accretion-dominated Ly$\alpha$ luminosity and the inner disk dust clearing, painting a picture where water molecules in the inner 3 au are exposed to and dissociated by strong Ly$\alpha$ emission as the opacity of the inner disk declines with time.  

\end{abstract}



\section{INTRODUCTION}
\label{sec:intro}

The circumstellar environments in which planets form are dynamic and volatile places.   Mass accretion onto the central stars is active and rapid stellar rotation drives strong magnetic fields. These processes power high levels of UV and X-ray emission, as well as temporal variability that alters the physical state and longevity of the protoplanetary disk.  The stellar high-energy radiation (UV and X-ray) controls circumstellar disk dispersal~\citep{alexander14,pascucci22}, strongly influencing the final stages of planet formation.  
 Young stars complete their mass assembly at the same time as protoplanets are forming around these objects  (1~--~10 Myr; \citealt{haisch01,fedele10, ribas14}),  therefore, there is strong competition between dynamical disk clearing through forming protoplanets (e.g., \citealt{rice03,dodson11,vandermarel18}) and disk dispersal through photoevaporation and disk winds, which depend on the strength and shape of the pre-main sequence far-ultraviolet (FUV; 912~--~1800~\AA) radiation field 
 \citep{alexander14,ercolano16,ercolano17,wang_goodman17}. The high-energy radiation produced in the accretion region and surrounding shock-heated gas is responsible for the high-levels of UV continuum and line radiation from Classical T Tauri Stars (CTTSs; \citealt{ingleby11,france14,schneider20}). 

\begin{figure*}[htpb]
   \centering
   \includegraphics[scale=.60,clip,trim=0mm 0mm 0mm 0mm,angle=0]{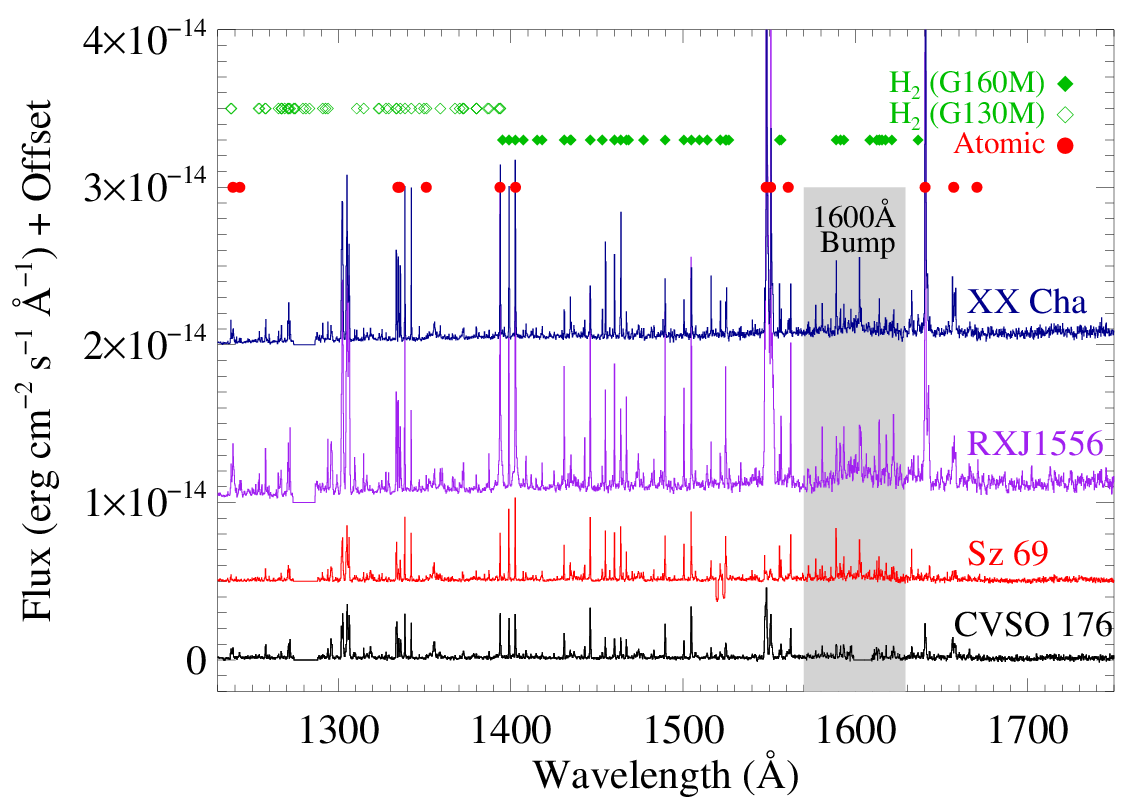}

   \figcaption{Example $HST$-COS spectra of four representative CTTSs from the ULLYSES survey.  The stellar spectra are color-coded and labeled in the legend in the lower left; offsets were applied to Sz 69 (5~$\times$~10$^{-15}$ erg cm$^{-2}$ s$^{-1}$ \AA$^{-1}$), RXJ1556.1-3655 (the flux was divided by 2 and an offset of 5~$\times$~10$^{-15}$ erg cm$^{-2}$ s$^{-1}$ \AA$^{-1}$ was applied), and XX Cha (20~$\times$~10$^{-15}$ erg cm$^{-2}$ s$^{-1}$ \AA$^{-1}$).  A 3 resolution element (21 pixel) boxcar smooth has been applied for plotting the full spectral range.  The wavelengths of prominent emission lines of H$_{2}$ (green diamonds, separated by lines that are primarily recorded in the G130M or G160M modes) and atomic lines (red circles) are indicated.  The peak wavelength region of the 1600~\AA\ Bump emission from Ly$\alpha$ dissociated H$_{2}$ (originating from a Ly$\alpha$ dissociated H$_{2}$O parent population) is indicated by the gray shaded region.  The $HST$-COS observations of CVSO 176 used a single central wavelength setting which results in a~$\approx$~12~--~15~\AA\ gap in the spectra where the COS detector segments a and b meet.
   }

\end{figure*}

 Current studies of the accretion of circumstellar disk material onto young stars explore the competing processes of magnetorotational instability (MRI) and magnetohydrodynamic (MHD) disk winds for removing excess angular momentum of from the disk and allowing material to accrete onto the central protostar.  In the MRI scenario, gas within $\sim$~0.1~au is ionized by the central star and entrained in the stellar magnetic field~\citep{balbus91}, while the MHD disk wind picture proposes that outflowing gas from the disk surface at radii out to tens of au (see, e.g., \citealt{lesur21} for a recent review) removes angular momentum from the disk.  Ultimately, material is funneled onto the central star along field lines where it creates an accretion shock at (possibly) several contact points with the pre-main sequence photosphere~\citep{calvet98,muzerolle01,hartmann16,pittman22}. 
  The final phase of mass accretion (stellar ages 2~--~10 Myr) is particularly important as accretion luminosity and stellar activity catalyze the dispersal and chemistry of the protoplanetary disk, impacting the timescales for planet formation and the accretion of protoplanetary atmospheres \citep{wang_goodman17}.     


The high-energy radiation from the accretion region is also responsible for ionizing and photodissociating inner disk atoms and molecules, catalyzing disk surface chemistry, and driving many of the observable signatures of the inner regions of the disk.   Molecular gas emission and absorption originating inside of 10 au provide our best means of estimating the conditions at the radii where gas giant and rocky planet cores are forming and accreting their nascent atmospheres. Surveys of molecular emission from the inner few au provide constraints on the radial distribution, temperature, and composition of planet-forming inner disks.  Surveys of mid-IR emission from CO~\citep{salyk09, brown13, banzatti15, banzatti22}, H$_{2}$O and organic molecules~\citep{pontoppidan10b,carr11, salyk11b,banzatti23}, UV emission and absorption of H$_{2}$ and CO~\citep{herczeg04, france11b, schindhelm12a, france14b, arulanantham21}, and spectrally/spatially-resolved near-IR observations~\citep{pontoppidan11,carmona11,brittain15} have placed constraints on the relative abundance ratios of H$_{2}$, CO, and H$_{2}$O, the evolution of the inner gas disk radius, and the excitation conditions of the molecular gas.  The inventory of disk-molecules and photochemical studies of the planet-forming environments around young stars is expected to expand rapidly in the era of $JWST$~\citep{oberg21,grant22,kospal23}.


H$_{2}$ is the primary mass component of gas-rich disks, however, the lack of intrinsic dipole moment and large energy spacing between the rotational levels have traditionally made direct detection of the rovibrational emission lines of the molecule challenging to study in a large number of disks (although see, e.g., \citealt{carmona11,gangi20, kospal23}).  Conversely, the electronic excitation spectrum of H$_{2}$ is among the brightest features in UV spectra of pre-main-sequence stars~\citep{france12}, and can serve as a diagnostic of weak mass accretion when traditional accretion diagnostics (e.g., H$\alpha$, NUV Balmer continuum) do not provide clear evidence of active accretion~\citep{ingleby11,alcala19}.

The molecular tracers observed at UV wavelengths are primarily excited by UV fluorescence and dissociation.  Fluorescent H$_{2}$~\citep{herczeg02} emission is detected in the circumstellar environments of all accreting protostars surveyed to date (e.g., \citealt{france12}) and is a sensitive probe of remnant gas in disks with evolved dust opacity~\citep{hoadley15}.  Fluorescent CO emission is observed in a fraction of disks, with an observed bias towards disks that show dust clearing in their mid-IR spectral slopes~\citep{schindhelm12a,arulanantham21}.  H$_{2}$ dissociation continuum emission (the ``1600~\AA\ Bump''), where Ly$\alpha$ photons dissociate water and the resultant, highly non-thermal H$_{2}$ dissociation products, is observed in $\sim$all dust-evolved disks~\citep{france17}, however, only approximately a third of ``full'' or ``primordial'' disks display this feature.  Together, the fluorescent H$_{2}$ line emission and continuum bump trace both the radial distributions and UV irradiation of hot molecular gas in the inner disks.

In this paper, we present a new survey of the H$_{2}$ features observed in the UV spectra of CTTSs, taking advantage of the Ultraviolet Legacy Library of Young Stars as Essential Standards (ULLYSES) program.  ULLYSES obtained almost 500 orbits of ultraviolet spectroscopic observations with the {\it Hubble Space Telescope} ($HST$), targeting approximately 71 new accreting and non-accreting young stars.  We analyze the ULLYSES Data Release 5 (DR5) to provide the largest spectral catalog of H$_{2}$ line and continuum features obtained to date to better understand the radial distribution and evolution of the inner disk molecules around young stars.

This paper is laid out as follows:  Section 2 presents an overview of the ULLYSES FUV spectra analyzed here and Section 3 presents the emission line measurements and a publicly-available data catalog of the FUV line parameters.  Section 3 concludes with the derivation of the Ly$\alpha$ luminosity, average H$_{2}$ fluorescent emitting radius, and the properties of the FUV continuum radiation.  In Section 4, we discuss how these results suggest a physical picture where stellar/accretion photons have greater penetration depth and drive greater molecular dissociation as the dust opacity of the inner disk declines.  We conclude with a brief summary in Section 5.

\begin{figure}[htbp]
   \centering
   \includegraphics[scale=.4,clip,trim=0mm 0mm 5mm 5mm,angle=270]{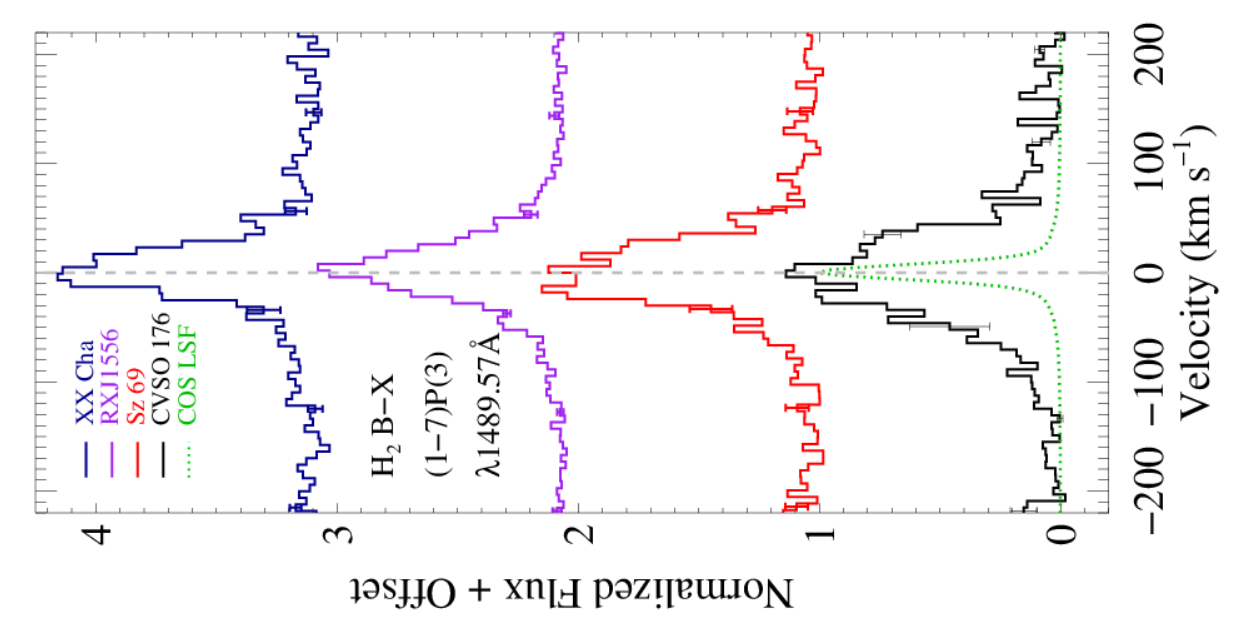}
   \figcaption{Individual H$_{2}$ $B$$^{1}\Sigma^{+}_{u}$~--~$X$$^{1}\Sigma^{+}_{g}$ (1~--~7) R(3) ($\lambda_{lab}$~=~1489.57~\AA) emission lines for the four example targets from Figure 1.  The spectra are shifted to their rest wavelength for comparison and normalized by the average flux in the peak ($\pm$~10 km s$^{-1}$) of the emission line and offset by 1.0.  The $HST$-COS line-spread-function (LSF) is shown as the dotted green line; all emission lines are well-resolved in the COS data, indicating kinematic line-broadening.   }
\end{figure}

\section{Archival Far-Ultraviolet Spectra from ULLYSES}
\label{sec:obs}

We analyze FUV spectra obtained as part of the 487 $HST$ orbits in the low-mass star portion of ULLYSES (see ~\citealt{duval20} for a program overview).  The observations include data acquired through Data Release 5\footnote{https://ullyses.stsci.edu/ullyses-download.html} with the addition of five additional targets observed in Summer 2022.  The final sample (Table 1) is primarily composed of accreting T Tauri stars observed as part of the ULLYSES program, however, the DR5 catalog includes a number of stars from previous young star observing programs with similar observational plans (see, e.g.,~\citealt{ardila13,france17}).  The final source list analyzed here includes 63 accreting and 8 non-accreting sources.  

 The target sample presented in ULLYSES DR5 were obtained between late-2009 and mid-2022 with the $HST$-Cosmic Origins Spectrograph (\citealt{green12}; COS).   As the data collection strategy and analysis of FUV spectra of accreting protostars has been discussed extensively in previous works~\citep{duval20,france12}, we present a high-level overview of the dataset here with an emphasis on the FUV spectral line catalog creation in Section 3.  The FUV spectral reduction and analysis follow the general description from the ODYSSEUS program overview paper~\citep{espaillat22}, and we refer the reader to that work for an example of the FUV spectra analyzed in the context of the larger ODYSSEUS panchromatic analysis program.  We also employ mid-IR spectral slopes (12~--~22$\mu$m) from $WISE$ photometry~\citep{cutri13}.  The $WISE$ data was downloaded from the NASA IPAC Infrared Science Archive and $WISE$ observations with data quality flags ({\it cc\_{flags}}~$\neq$~0) were excluded from the analysis.   In addition, visual inspection of the $WISE$ images indicated potential field crowding or extended emission in CVSO-90, CVSO-107, CVSO-176, RECX-15 (ET Cha, echa j0843.3-7915), and CS Cha.  Of these, we note that CS Cha hosts a well-studied transition disk~\citep{espaillat07a} providing confidence in the large $WISE$ W3~--~W4 color for this source.    

The $HST$ observations were made using the COS G130M and G160M grating modes.  Most observing programs employed multiple central wavelength settings at several focal-plane split positions to create continuous FUV spectra from $\sim$ 1140~--~1760~\AA\ and mitigate the effects of fixed pattern noise\footnote{We note that some of the ULLYSES observations used a single central wavelength setting which results in ~$\approx$~12~--~15~\AA\ gaps in the spectra where the COS detector segments a and b meet.}.   These modes provide a point-source resolution between $\Delta$ $v$~$\approx$~17~--~22 km s$^{-1}$ with 6~--~7 pixels per resolution element~\citep{osterman11}. 
The total FUV exposure times were between two and sixteen orbits per target, depending on the intrinsic luminosity and the interstellar plus circumstellar reddening on the sightline. 

We present an example of four representative T Tauri star spectra from ULLYSES in Figure 1 (XX Cha, RXJ1556.1-3655, Sz 69, and CVSO 176).   These stars come from a variety of star-forming regions sampled by ULLYSES (Chamaeleon I, Corona Australis, Lupus, and Orion OB1) and the locations of prominent molecular and atomic features in the spectra are shown.  As this work focuses primarily on the line and continuum emission from H$_{2}$, Figures 2 and 3 present representative samples of these features.  Figure 2 shows a representative, isolated, fluorescent emission line of H$_{2}$ ($B$$^{1}\Sigma^{+}_{u}$~--~$X$$^{1}\Sigma^{+}_{g}$ (1~--~7) R(3); $\lambda_{lab}$~=~1489.57~\AA) for the four stars shown in Figure 1.  Figure 3 presents the range of H$_{2}$ continuum spectra observed in ULLYSES DR5:  the top panel shows the underlying FUV spectral continuum (CVSO-109) showing no molecular emission (see also~\citealt{espaillat22}) while the middle panel shows the same spectral extraction from a source with a strong molecular continuum feature (``1600~\AA\ Bump'', \citealt{france17}; DM Tau).  The bottom panel of Figure 3 shows the extracted H$_{2}$ molecular continuum from DM Tau (see Section 3.3).

\begin{figure}[htbp]
   \centering
   \includegraphics[scale=.55,clip,trim=0mm 0mm 0mm 0mm,angle=0]{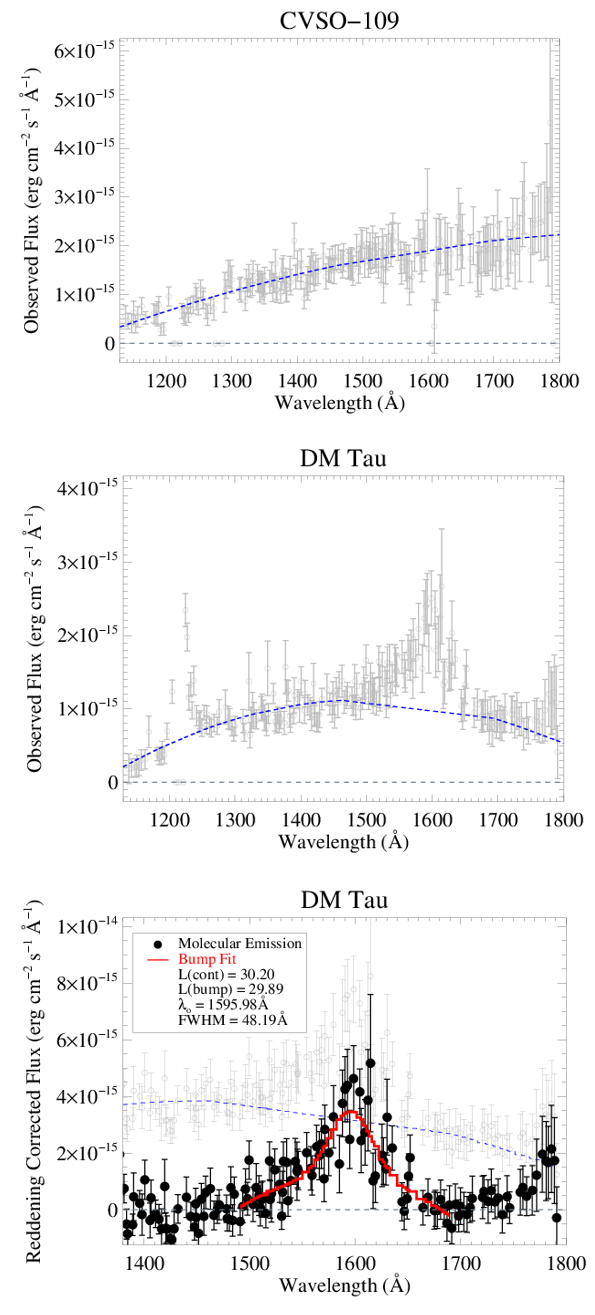}
   \figcaption{Example FUV continuum spectra for two targets in the ULLYSES DR5, CVSO 109 (top panel; \citealt{espaillat22}) and DM Tau (center panel).  The underlying accretion continuum (fit with a polynomial; blue dashed line) is removed and the residual molecular continuum is shown as the black circle data points in the bottom panel.   The molecular continuum fit is parameterized by a Gaussian + polynomial~\citep{france17} and is shown in red.  Molecular continuum fit parameters are shown in the legend;  the log$_{10}$(luminosity) of the FUV continuum and the 1600~\AA\ Bump, in units of erg s$^{-1}$, are presented as L(cont) and L(bump), respectively}.  
\end{figure}

\section{FUV Spectral Line Data Reduction and Analysis}

We analyzed the brightest FUV lines in the COS spectra: five progressions of fluorescent H$_{2}$ ([1,7], [1,4], [0,1], [0,2], and [2,12]) and hot gas emission from \ion{Si}{3} (1206~\AA), \ion{N}{5} (1238~\AA\ and 1243~\AA), \ion{Si}{4} (1394\AA\ and 1403\AA), and \ion{C}{4} (1548\AA\ and 1550\AA).   The H$_{2}$ lines are marked in Figure 1, depending on whether they were recorded on the G130M (open diamonds) or G160M (filled diamonds) grating and the hot gas lines are marked with red circles.  The spectra and tables of line flux measurements are publicly available in machine-readable format\footnote{{\tt https://lasp.colorado.edu/home/cusp/outflows-and-disks/}}.  For each star, we take the high-level processed data from the ULLYSES *cspec.fits files and extract individual emission line flux and error regions, which are saved as line-specific arrays.    
For hot gas doublets (e.g., \ion{C}{4}, \ion{N}{5}, and \ion{Si}{4}), the lines are extracted separately into the blue and red doublet components.  Individual extracted emission line region and tables of line measurements for the ensemble of stars are available on the web portal.  

Each array is processed with a 5-pixel boxcar smoothing algorithm to suppress noise without compromising significant kinematic information (all of the emission lines studied here are well-resolved see, e.g., \citealt{france12b}; Figure 2).  The $HST$-COS line spread function contributes $\sim$20\% for lines narrower than 40 km s$^{-1}$ and less for broader lines.  For each emission line region, background continuum flux is calculated from an immediately adjacent spectral region then subtracted out from the entire flux array.  In Table 2, we present the integrated fluxes measured over the interval $\pm$~200 km s$^{-1}$. 

The full width at half maximum (FWHM) was calculated non-parametrically to account for the various non-Gaussian shapes of some of the emission lines~\citep{ardila13}. The brightest 7 pixels (approximately one resolution element) in an emission line region are averaged to find the maximum flux, F$_{max}$. The line center is defined as the average wavelength within the 7 pixel F$_{max}$ region.  The half maximum value, $F_{max/2}$ is calculated as 
\begin{equation}
    F_{max/2} = (F_{max}-F_{cont})/2  + F_{cont} 
\end{equation} 
where $F_{cont}$ is the value of the continuum emission near the line of interest.  The $\pm$~half-width points are the wavelengths corresponding to F$_{max/2}$; the difference of these half-width points are the FWHMs presented in Table 2. For the hot gas doublets the FWHM was only calculated on the blue component of the doublet because these doublets are governed by the same physical creation mechanism and the blue components of the doublet are stronger (brighter).  Uncertainties on the FWHM are computed using the same method, with the $\pm$1 $\sigma$ flux error added to the flux values to evaluate the minimum and maximum values achievable within the photometric flux uncertainty.    

In certain instances (approximately 5\% of line profile measurements), the FWHM calculation returned values that did not match a qualitative inspection of the data, and for these lines, we measured the line profiles following the method described by~\citet{france12b}, where a Gaussian line shape is forward-modeled and compared to the data using a  Levenberg-Marquardt least-squares minimization to return the best-fit emission line parameters.  This “patch” was tested against approximately a dozen emission lines with good non-parametric results, and the agreement in FWHM and flux was found to be within 10-20\% in all cases.  
	The emission line measurement files are stored as comma separated value arrays and are publicly available~\footnote{ {\tt https://lasp.colorado.edu/home/cusp/outflows-and-disks/ } }. Each data file includes the emission line IDs, the peak wavelength of each line, the integrated flux of the line, the error on that integrated flux, the FWHM of the line, the error on that FWHM and a data quality flag. The data quality flags note special circumstances or indicate when there is no emission line. In the following subsections, we describe how these basic data products were used to derive system parameters of the ULLYSES targets (\S3.1 and \S3.2) and how the FUV continua were extracted from the $HST$-COS datasets (\S3.3).

\subsection{H$_{2}$ Progression and Ly$\alpha$ Luminosity from the ULLYSES Sample}

The fluorescent H$_{2}$ lines observed in the CTTS sample can be used to constrain the spatial distribution of H$_{2}$ in the circumstellar environment. For our line-profile analysis, we focus on the measurement of four progressions ([$v^{'}$,$J^{'}$] = [1,7], [1,4], [0,1], and [0,2]). These progressions are pumped through the (1~--~2)R(6) $\lambda_{lab}$1215.73~\AA, (1~--~2)P(5) $\lambda_{lab}$1216.07~\AA, (0~--~2)R(0) $\lambda_{lab}$1217.21~\AA, and (0~--~2)R(1) $\lambda_{lab}$1217.64~\AA\ absorbing transitions, respectively\footnote{The quantum numbers $v$ and $J$ denote the vibrational and rotational quantum numbers in the ground electronic state ($X$$^{1}\Sigma^{+}_{g}$), the numbers $v^{'}$ and $J^{'}$ characterize the H$_{2}$ in the excited electronic state ($B$$^{1}\Sigma^{+}_{u}$), and the numbers $v^{''}$ and $J^{''}$ are the rovibrational levels in the electronic ground state following the fluorescent emission. Absorption lines are described by ($v^{'}$~--~$v$) and emission lines by ($v^{'}$~--~$v^{''}$).}. The absorbing transitions are within +14~--~+487 km s$^{-1}$ of Ly$\alpha$ line-center.  We target the brightest two, non-blended emission lines from each progression for analysis here.  The H$_{2}$ emission line centroids are broadly consistent with the stellar radial velocities within the 15 km s$^{-1}$ accuracy of the $HST$-COS wavelength solution, and we will explore multiple emission line components (e.g., \citealt{arulanantham18}) and their kinematic signatures in a future work.

The emission line fluxes are corrected for interstellar reddening using the A$_{V}$ values given in Table 1.  It has been shown that the dust properties of the Taurus star-forming region differ from the Milky Way average extinction curve~\citep{calvet04}.  However, the shape of the extinction curve for star-forming regions and individual targets remains highly uncertain~\citep{mcjunkin16}, and the A$_{V}$ values derived for individual target carry significant uncertainties~\citep{pittman22,carvalho22}.  Both of these issues are exacerbated by the strong wavelength-dependence of the dust attenuation curve, where the reddening correction at 1500~\AA\ has a $\sim$~6x larger amplitude relative to 6500~\AA\ for A$_{V}$ = 1.  In light of these uncertainties, we adopt a typical interstellar curve~\citet{ccm89} and an $R_{V}$~= 3.1.  If, on the other hand, we would have adopted a standard interstellar curve with $R_{V}$~= 5.5 (for A$_{V}$ = 1), the resultant fluxes would be between 20~--~40\% of the values reported here.

The total flux from a progression $m$ is given by 
\begin{equation}
F_{m}(\textup{H$_{2}$}) = \frac{1}{2} \sum \left ( \frac{F_{mn}}{B_{mn}} \right ) 
\end{equation}
where $F_{mn}$ is the reddening corrected, integrated H$_{2}$ emission line flux 
from rovibrational state $m$ (= [$v^{'}$,$J^{'}$]) in the $B$$^{1}\Sigma^{+}_{u}$ electronic state to $n$ (= [$v^{''}$,$J^{''}$]) in the ground electronic state, $X$$^{1}\Sigma^{+}_{g}$. $B_{mn}$ is the branching ratio between these two states, and 2 is the number of emission lines measured from a given progression. 
The measurement errors are typically small, so we take the flux error to be the average error of the individual measurements of $F_{m}$(H$_{2}$).  The dominant systematic error on the measured H$_{2}$ flux is the correction for interstellar reddening; we do not attempt to account for this uncertainty in the flux and luminosity errors presented below. The total progression luminosity is then $L_{m}$(H$_{2}$)~=~(4$\pi$$d^{2}$)$F_{m}$(H$_{2}$). In Table 2, we present reddening-corrected luminosities for the [1,4] and [0,1] progressions. The total H$_{2}$ luminosity is the sum of the individual progression luminosities.  Because we select a finite number of progressions to analyze, this treatment underestimates the total H$_{2}$ fluorescent luminosity by an amount that depends on the temperature and spatial distribution of the absorbing H$_{2}$, as well as the shape of the Ly$\alpha$ pumping spectrum.  As these four progressions are the brightest in essentially all CTTSs, we consider this to be a 20~--~30\% underestimate of the total fluorescent H$_{2}$ emission~\citep{france12}.

\begin{figure}[htbp]
   \centering
   \includegraphics[scale=.46,clip,trim=0mm 0mm 0mm 0mm,angle=0]{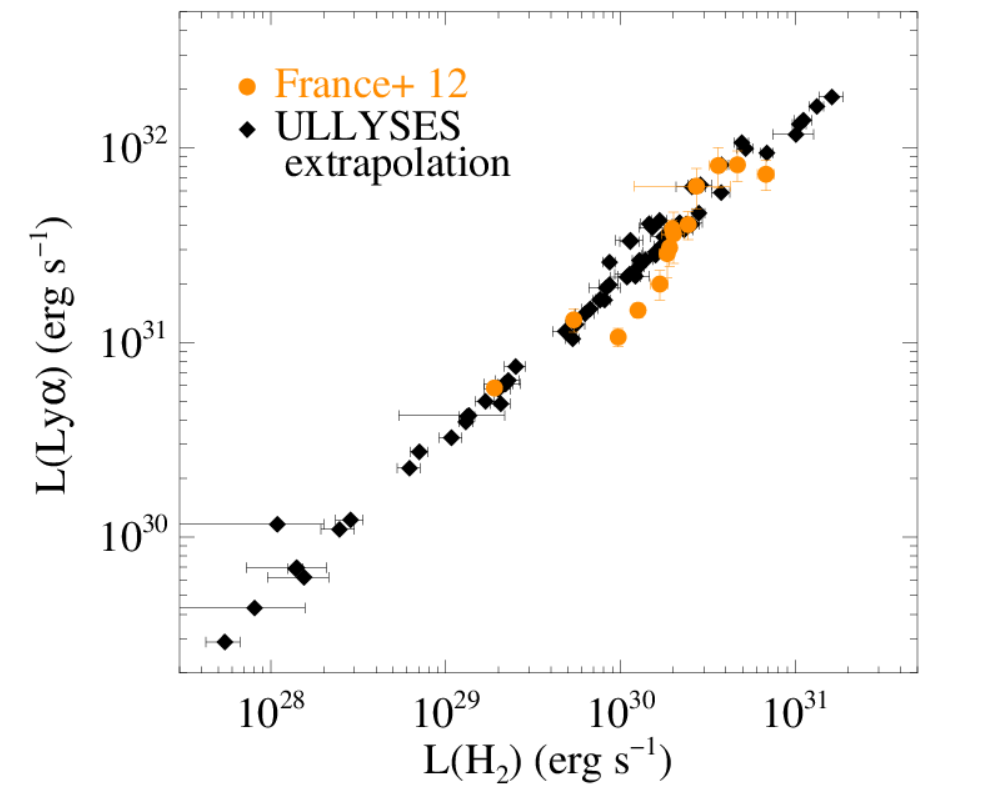}
   \figcaption{Projected Ly$\alpha$ luminosities for all ULLYSES DR5 stars (black diamonds) as a function of total fluorescent H$_{2}$ luminosity. We created a power-law fit to the 14 stars with high-quality, multi-progression H$_{2}$ emission line data and Ly$\alpha$ reconstructions from \citet{france12}, shown as orange circles on the plot.  This fit was used to predict the Ly$\alpha$ luminosity of the ULLYSES stars based on their observed F(H$_{2}$).   }
\end{figure}

\begin{figure}[htbp]
   \centering
   \includegraphics[scale=.48,clip,trim=0mm 0mm 5mm 5mm,angle=0]{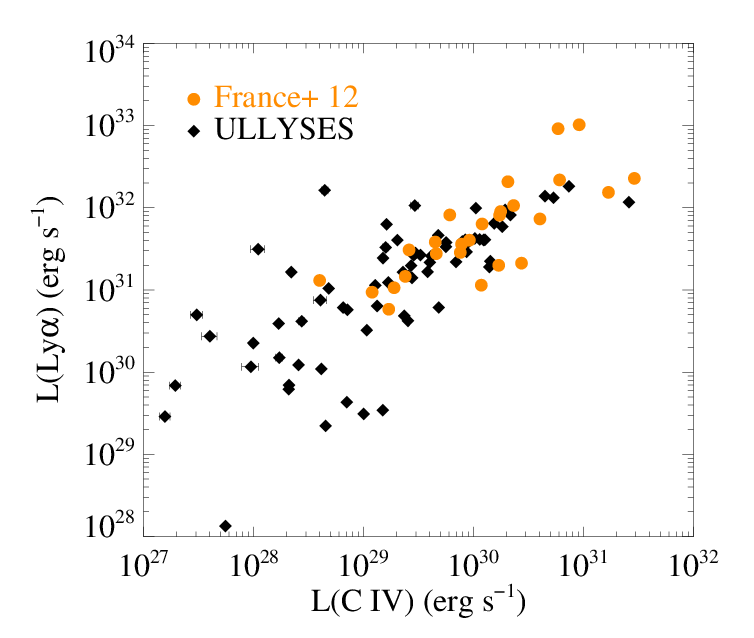}
   \figcaption{Comparison of the \ion{C}{4} and Ly$\alpha$ luminosities of the ULLYSES stars (black diamonds) and previous $HST$-COS survey of CTTS (\citealt{france12b}; orange circles). The plot demonstrates the correlation between Ly$\alpha$ (as derived in Section 3.1) and accretion-dominated emission from \ion{C}{4}.  This is consistent with the broad, bright Ly$\alpha$ profiles of CTTS~\citep{schindhelm12b,arulanantham23} being powered by mass accretion (e.g., \citealt{france14}).  }
\end{figure}

The fluorescent emission lines of H$_{2}$ can be used to reconstruct the bright Ly$\alpha$ emission line that is the primary excitation source of this gas~\citep{wood02,herczeg04,schindhelm12b}.  In order for this process to produce a high-fidelity reconstruction, a sufficient number of H$_{2}$ progressions, covering the majority of the Ly$\alpha$ pumping profile, must be detected at moderate to high S/N~(e.g., \citealt{schindhelm12b,arulanantham18}).   Ly$\alpha$ reconstructions with a small number of low-S/N progressions leave the shape and integrated flux of the Ly$\alpha$ line poorly constrained~\citep{espaillat22}.  Of the 63 accreting sources analyzed here, the majority are only detected in 3 or 4 H$_{2}$ progressions~--~54 detections in [1,4], 41 detections in each [1,7] and [0,1], 33 detections in [0,2], and only 3 detections in the [2,12] progression that is often used to constrain the wings of the Ly$\alpha$ line.   Therefore, a full Ly$\alpha$ reconstruction cannot be practically applied to the full ULLYSES sample.    

In order to provide a uniform estimate for the Ly$\alpha$ luminosity for ULLYSES, we fit F(H$_{2}$) vs. F(Ly$\alpha$) relation using the 14 higher S/N targets reconstructed by~\citet{schindhelm12b} and H$_{2}$ fluorescence data from~\citet{france12}.   These data are shown as the orange points in Figure 4.  We extrapolated the ULLYSES measurements of L(H$_{2}$) described above onto this relation to estimate the Ly$\alpha$ fluxes for all 63 accreting stars in our sample:  
\begin{equation}
log_{10}(F(Ly\alpha)) = m_{Ly\alpha}(log_{10}(F(H_{2}))) + b_{Ly\alpha}
\end{equation}
where the best-fit coefficients are $m_{Ly\alpha}$ = 0.791~$\pm$~0.033 and $b_{Ly\alpha}$ = $-$1.297~$\pm$~0.400.
Propagating the uncertainties on the fit coefficients of the F(H$_{2}$) vs. F(Ly$\alpha$) relation indicates a maximum uncertainty on the associated Ly$\alpha$ luminosities of $\lesssim$40\%.  The small dispersion seen in the extrapolated Ly$\alpha$ luminosities is due to the conversion of the flux-flux relationship from~\citet{france12} to luminosity. We caution that while the original Ly$\alpha$-to-H$_{2}$ relationship from~\citet{france12} spans approximately 1.5 orders of magnitude in L(H$_{2}$), we are extrapolating that relationship to over 3 orders of magnitude for the present sample.   Figure 5 illustrates the comparison of the estimated Ly$\alpha$ to \ion{C}{4} luminosity between the present work and the \citep{france12} sample, showing reasonable agreement in the range $L$(\ion{C}{4})~$>$~10$^{29}$ erg s$^{-1}$ with larger spread in the lower \ion{C}{4} luminosity region of the ULLYSES sample.  The spread in Ly$\alpha$ values at $L$(\ion{C}{4})~$<$~10$^{29}$ erg s$^{-1}$ is driven by scatter in the relationship between \ion{C}{4} and H$_{2}$.  Similarly, we a moderate positive correlation between L(Ly$\alpha$) and the photometric luminosity (L$_{*}$; from \citealt{manara21, manara22}): the Spearman rank correlation coefficient, $\rho$, and the likelihood that this correlation is consistent with a random distribution, n, for the L(Ly$\alpha$) vs. L$_{*}$ relation are $\rho$~= 0.498 and n = 2.3~$\times$~10$^{-4}$.


\subsection{Radial Distribution of H$_{2}$ Fluorescent Emission}

Kinematic broadening dominates the observed H$_{2}$ line profiles. The thermal broadening of the emission lines is approximately 4.5 km s$^{-1}$ at the nominal 2500 K H$_{2}$ layer~\citep{herczeg04,hoadley15}; significant additional broadening would require temperatures in excess of the $\approx$~4500 K dissociation temperature of H$_{2}$~\citep{lepp83}. If we further assume that any turbulence in the disks is subsonic, then the maximum turbulent velocity will be no larger than a few km s$^{-1}$. Therefore, velocity broadening due to bulk motions and Keplerian rotation dominate the observed line shapes when the FWHM of the emission line is greater than the $\approx$~15~--~20 km s$^{-1}$ spectral resolution of COS (see Figure 2; all H$_{2}$ lines are well-resolved at the spectral resolution of $HST$-COS; the 1450~\AA\ $HST$+COS LSF overplotted as the green dotted line). 
For the case of H$_{2}$ in a circumstellar disk where Keplerian broadening dominates the profile, we employ a simple metric to characterize the average H$_{2}$ emitting radius, $\langle$$R_{H2}$$\rangle$~\citep{salyk11,france12,arulanantham21}, where
\begin{equation}
\langle R_{H2} \rangle_{m} = GM_{*} \left (\frac{2 sin(i)}{FWHM_{m}} \right )^{2} 
\end{equation} 
where $M_{*}$ is the stellar mass, $i$ is the disk inclination angle, taken primarily from ALMA surveys (Table 1), and FWHM$_{m}$ is the mean of the Gaussian FWHMs for a given progression $m$. When disk inclinations were not available (19/63 accreting sources), we assume a disk inclination of 60\arcdeg, as this is approximately the average inclination found in ALMA disk surveys~\citep{ansdell18} and assumed in model studies~\citep{rilinger21}.  We provide correlation coefficients for the full sample and using only the sources with known inclinations in the appropriate subsections below, and note that the ``outlier'' values of $\langle$$R_{H2}$$\rangle$ are not driven by unknown inclinations; 5/6 of the largest H$_{2}$ radii presented here are for sources with measured disk inclinations. Table 1 presents the stellar masses and disk inclinations used in this work.   


For the H$_{2}$ radial distribution analysis, we consider four progressions:  the [1,7] and [1,4] progressions that are pumped within 100 km s$^{-1}$ of the Ly$\alpha$ line center and two progressions ([0,1] and [0,2]) that are pumped from the wing of the Ly$\alpha$ emission profile (379 and 487 km s$^{-1}$ from line-center, respectively). We analyze multiple progressions because self-absorption can selectively impact the lower-energy level gas~\citep{wood02,mcjunkin16} and that off-star Ly$\alpha$ emission and the local scattering geometry can impact the flux distributions of the different progressions~\citep{walter03,arulanantham23}.  
We will return to a discussion of these results in Section 4.

\subsection{H$_{2}$ Dissociation Emission and the FUV Continuum}

FUV continuum spectra were created for the ULLYSES DR5 sample following the continuum extraction methodology described in~\citet{france14a}.   We create a grid of 205 unique spectral points between 1138 and 1791 \AA, selected by hand to avoid discrete molecular and atomic emission and absorption features, where 0.75~\AA\ (approximately 10 spectral resolution elements) spectral continuum windows can be cleanly measured.  We measure the mean and standard deviation of the observed spectra, and these points define the binned flux spectrum and error array (see Figure 3).  These binned spectra are then corrected for interstellar reddening and the FUV continuum and H$_{2}$ continuum emission (the ``1600~\AA\ Bump'') are separated as described below.     

\begin{figure}[htbp]
   \centering
   \includegraphics[scale=.42,clip,trim=0mm 0mm 0mm 0mm,angle=0]{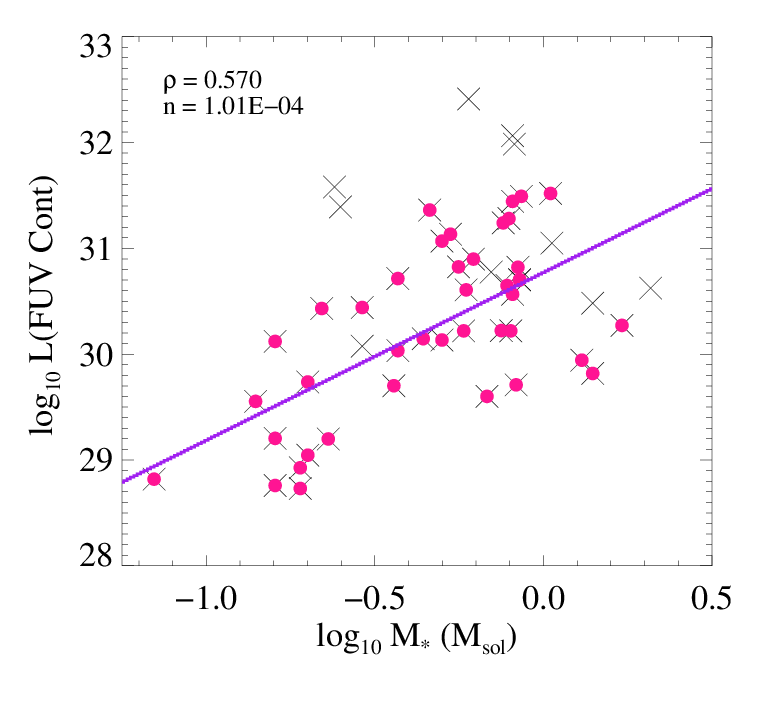}
   \figcaption{The FUV continuum luminosity increases with pre-main sequence stellar mass.  The stars in the ULLYSES sample with good measurements of the average H$_{2}$ emitting radius are shown with a black x, and those with literature A$_{V}$~$<$~1 are overlaid with pink circles (see Section 3.3 and 4.1).  A fit to the A$_{V}$~$<$~1 points is overplotted in purple.  The Spearman rank coefficient ($\rho$) and probability coefficient for null correlation (n) of the A$_{V}$~$<$~1 points are shown at the upper left.}
\end{figure}

We measure the FUV continuum emission by fitting the binned spectrum with a second order polynomial at wavelengths away from the Bump and residual stellar or airglow emission, in the regions $\Delta$$\lambda$~=~1145~--~1190~\AA, 1245~--~1330~\AA, 1395~--~1401~\AA, 1420~--~1465~\AA, 1690~--~1710~\AA, and 1730~--~1750~\AA\ (see the blue dashed lines in Figure 3).  The polynomial fit is extrapolated to the atomic hydrogen ionization limit, 912~\AA, to cover the primary photodissociating spectral region for H$_{2}$, CO, and N$_{2}$.  Comparison with archival $FUSE$ observations between 1000 and 1180~\AA\ has shown that this extrapolation is a reasonably good match to the observed short-wavelength flux~\citep{france14}.  The FUV continuum flux, $F_{FUV Cont}$, is the integral of the polynomial over the 912~--~1760~\AA\ wavelength region and the FUV continuum luminosity is defined as $L$(FUV Cont)~=~4$\pi$$d^{2}$$F_{FUV Cont}$.  

\begin{figure}[htbp]
   \centering
   \includegraphics[scale=.48,clip,trim=0mm 0mm 5mm 5mm,angle=0]{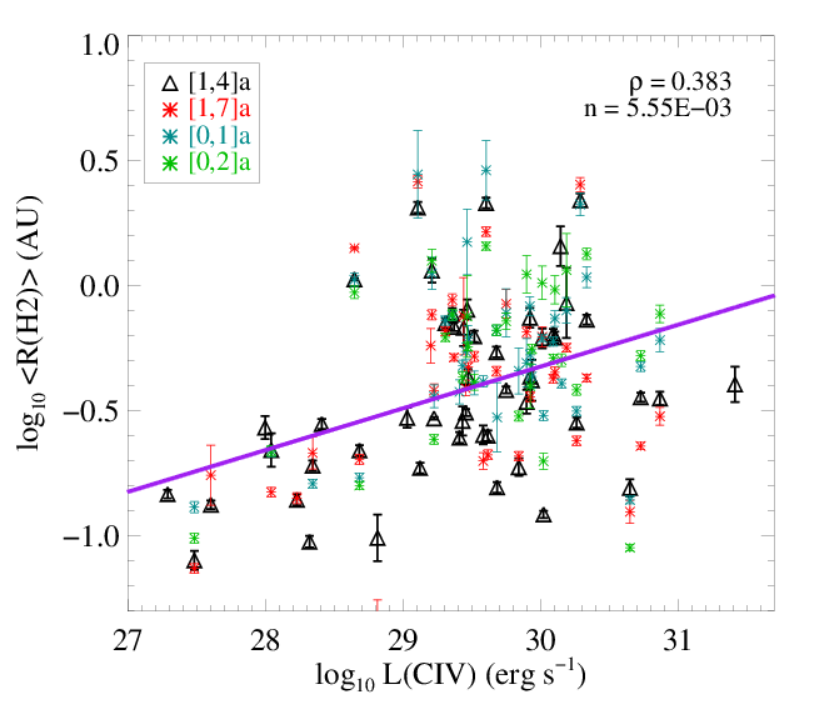}
   \figcaption{A comparison of the C IV luminosity with the average H$_{2}$ emitting radius for four of the progressions studied in this work (color-coded as indicated in the legend).  As has been shown in previous work, the majority of H$_{2}$ emission in CTTS systems arises between 0.1 and 3 au.  A best fit power-law relationship between L(\ion{C}{4}) and $\langle$$R_{H2}$$\rangle$ for the [1,4] progression is shown overplotted in purple and the Spearman rank coefficient ($\rho$) and probability coefficient for null correlation (n) are shown at the upper right. Assuming that \ion{C}{4} serves as a proxy for the accretion luminosity, the H$_{2}$ radial distribution demonstrates a weak dependence on the strength of the high-energy emission, possibly suggesting increased molecular dissociation in the innermost region of rapidly accreting systems.   }
\end{figure}

Figure 6 shows the FUV continuum luminosity as a function of stellar mass.   We observe a positive correlation between the continuum and stellar mass ($\rho$~= 0.570, n = 1.0~$\times$~10$^{-4}$ and shown in the legend of Figure 6), which is expected in the scenario where the FUV continuum emission is dominated by mass accretion onto the central star~\citep{france14a}.  Figure 6 plots the sources for which ``good'' H$_{2}$ radii are measured (see the discussion in Section 4.1), and it is observed that the most luminous sources are also the sources with the largest reddening (A$_{V}$).  The attenuation towards  FUV emitting regions of protoplanetary targets has been estimated to be lower than estimates inferred from X-ray, optical, or IR measurements~(e.g., \citealt{mcjunkin14}).  Hence, extinction values derived from, e.g., IR observations may overestimate the extinction correction for UV data.  This was demonstrated by~\citet{france17}, who showed that the largest A$_{V}$-values derived from IR-based extinctions led to unphysical correlations in FUV luminosity relationships.  
To avoid biasing the results by the largest, possibly incorrect, A$_{V}$ values, Figure 6 also shows the 41 stars with A$_{V}$~$<$~1 as pink circles.  The correlation coefficient and best-fit line are presented for the A$_{V}$~$<$~1 sample.   

The most prominent spectral feature in the binned spectra of some CTTSs is the 1600~\AA\ Bump feature created during the dissociation of H$_{2}$ molecules with highly non-thermal rovibrational distributions, which spans $\sim$~30~--~150~\AA\ and has a peak flux $\sim$~0~--~4 times the underlying continuum level at 1600~\AA.  We define the 1600~\AA\ Bump as the excess emission above the FUV continuum in the 1490~--~1690~\AA\ spectral region, where the Bump spectrum is the binned data minus the FUV continuum fit.  An example of the well-defined Bump spectrum of DM Tau is shown in Figure 3.  Following previous work, we parameterize the Bump emission as a broad Gaussian plus a second-order polynomial~(\citealt{france17}, Figure 3, {\it bottom panel}).  We define the total Bump flux,  $F_{Bump}$, as the integral of the continuum-subtracted Bump spectrum from 1490 to 1690~\AA, and $L$(1600~\AA\ Bump)~=~4$\pi$$d^{2}$$F_{Bump}$.  

\begin{figure}[htbp]
   \centering
   \includegraphics[scale=.44,clip,trim=0mm 0mm 5mm 5mm,angle=0]{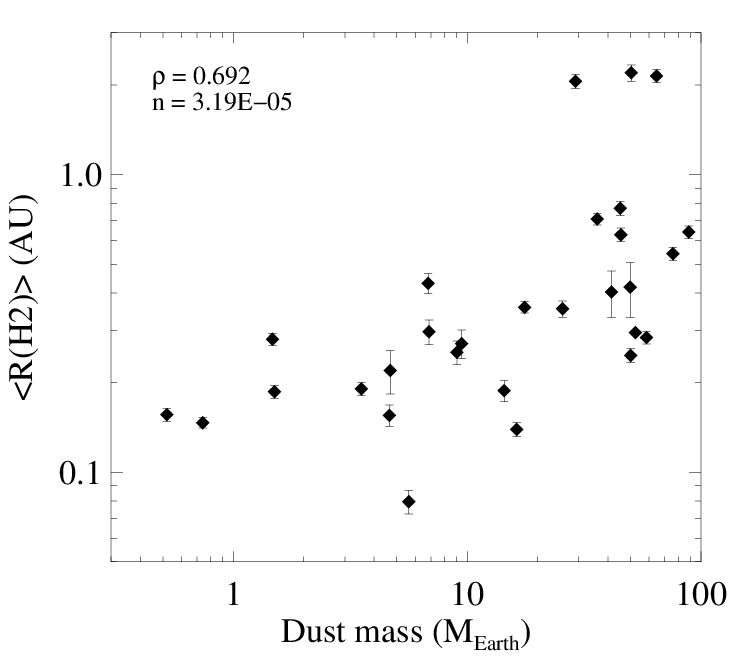}
   \figcaption{A comparison of the dust masses derived from mm-wave observations (see \citealt{manara22} for a review) with the average H$_{2}$ emitting radius for the [1,4] progression (29 sources had both dust masses, disk inclinations $>$~15\arcdeg, and good H$_{2}$ FWHM observations). The Spearman rank coefficient ($\rho$) and probability coefficient for null correlation (n) are shown at the upper left.  The behavior is consistent with a picture where more massive stars hosting more massive dust disks also photodissociate their inner few 0.1 au, leading to a correlation between disk mass and H$_{2}$ emitting radii.  }
\end{figure}

\begin{figure}[htbp]
   \centering
   \includegraphics[scale=.36,clip,trim=0mm 0mm 0mm 0mm,angle=0]{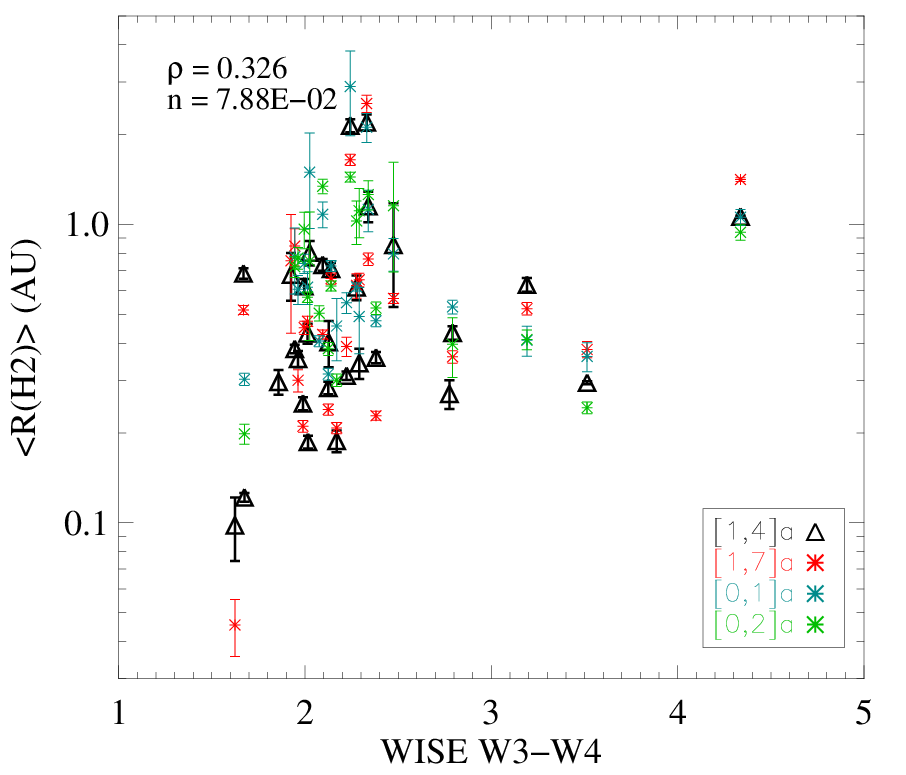}
   \figcaption{A comparison of the $WISE$ 12$\mu$m~-~22$\mu$m color with the average H$_{2}$ emitting radius for four of the progressions studied in this work (color-coded as indicated in the legend).  The distribution of emitting radii are clustered between W3 - W4~$\approx$~2.0 - 2.5 with a small number of transition disks (Sz 111, DM Tau, CS Cha) extending to large MIR flux ratios.  The Spearman rank coefficient ($\rho$) and probability coefficient for null correlation (n) are shown at the upper left.    }
\end{figure}

\section{Discussion:  The Spatial Distribution and Excitation Conditions of H$_{2}$ Line and Continuum Emission}

\subsection{Variation of $\langle$$R_{H2}$$\rangle$ with FUV Emission and Dust Environments}

The average H$_{2}$ line emission radius represents the characteristic  Keplerian velocity of the gas, assuming the spectral broadening is dominated by the rotational motion of the disk.  In a future work, we will present a more detailed line profile decomposition of a subset of bright disks to constrain stratified gas structures and wind contributions (e.g., \citealt{herczeg06,gangi23}), but for the full ULLYSES sample, we assume a single emission line component.  The objects included in the analysis are restricted to disk inclinations $>$~15 degrees (the Doppler motion cannot be confidently measured for face-on disks at the resolution of $HST$-COS) and do not have any data quality flags (low S/N, spectral blending) in the H$_{2}$ emission line catalog.  

The final sample of sources that meet these criteria (``good'' H$_{2}$ radii) includes 52, 38, 39, and 31 disks in the [1,4], [1,7], [0,1], and [0,2] progressions, respectively.  $\langle$$R_{H2}$$\rangle$ values range between 0.05~--~2.9 au, with average emitting radii between 0.53~--~0.72 au (median radii between 0.35~--~0.57 au) for the four progressions.  The majority of the fluorescent line emission (based on the standard deviation of the radial distributions of the four  progressions) arises between 0.1~--~1.4 au.  This typical emitting range can be seen in the cluster of radial distributions in Figure 7.  The minimum range of H$_{2}$ emitting radii (0.05~--~0.1 au) are comparable to the dust sublimation radius and the inner disk dust rim radii measured by GRAVITY (0.1~--~0.2 au; \citealt{gravity21}), suggesting that the molecular gas and dust edges may overlap in some sources.

%


Figure 7 shows the relationship between $\langle$$R_{H2}$$\rangle$ and the \ion{C}{4} luminosity ($\rho$~= 0.383, n = 5.6~$\times$~10$^{-3}$, reducing to $\rho$~= 0.330, n = 4.9~$\times$~10$^{-2}$ if restricted to disks with measured inclinations).  
Given the variability in mass accretion rates and FUV fluxes over time~\citep{espaillat19b,hinton22,claes22}, we use the $L$(\ion{C}{4}) measured contemporaneously with the H$_{2}$ emission lines to provide a proxy for the 'real-time' FUV radiation field of the central star.  In Figure 7, we observe a marginal positive correlation between the emitting radius and the strength of the \ion{C}{4} emission.  This result may indicate that higher FUV fluxes result in larger equilibrium radii with the heating/dissociating FUV radiation field (e.g., \citealt{cazzoletti18}), suggesting higher levels of H$_{2}$ dissociation, or the clearing of H$_{2}$-shielding inner disk dust, close to the central star in more FUV-luminous systems (Figure 6).  We note that the lowest \ion{C}{4} luminosity systems (log$_{10}$($L$(\ion{C}{4}))~$\lesssim$~28.7) populate the lower left quadrant in Figure 7 (log$_{10}$$\langle$$R_{H2}$$\rangle$~$\lesssim$~0.3).  Comparing the distribution of $L$(\ion{C}{4}) with the stellar mass reveals that all stars with log$_{10}$($L$(\ion{C}{4}))~$\lesssim$~28.7 also have M$_{*}$~$<$~0.3~M$_{\odot}$.  This supports the idea that the lowest mass sources with the lowest $L$(\ion{C}{4}) show less dissociation in their innermost disks than their more massive counterparts.  This would lead the stellar Ly$\alpha$ photons to encounter optically thick gas at smaller radial distances, which manifests as a small $\langle$$R_{H2}$$\rangle$ in these systems.  

We observe a stronger correlation ($\rho$~= 0.692, n = 3.2~$\times$~10$^{-5}$) between the average H$_{2}$ emitting radius and the disk dust mass (Figure 8).  The actual distribution is not a straight linear trend, instead, the H$_{2}$ emitting radii are roughly constant between $\approx$ 0.1~--~0.4 au for dust masses $\lesssim$~20 M$_{Earth}$, with no H$_{2}$ radii larger than $\approx$~0.4 au.   For dust masses  $\gtrsim$~20 M$_{Earth}$, the H$_{2}$ radii range from 0.2~--~2~au.   It is established that more massive stars host massive dust disks~\citep{andrews13, pascucci16,ansdell17} and that dust disk mass is correlated with mass accretion rate~\citep{manara16,mulders17,manara22}. The observation of larger gas clearing radii around stars with larger dust masses (and higher FUV continuum luminosities) is consistent with this picture of inner disk structure,  
as is the scenario presented above where the stars with the lowest masses (and \ion{C}{4} luminosities) were found to host H$_{2}$ populations with the smallest emitting radii.   \nocite{andrews13,pascucci16,ansdell17,manara16,mulders17,manara22} 

Figure 9 displays the relationship between the $WISE$ W3~--~W4 color\footnote{WISE W3 is centered at 12~$\mu$m, full band range~$\sim$~7.5~--~16~$\mu$m; WISE W4  is centered at 22~$\mu$m, full band range~$\sim$~20~--~25~$\mu$m; \citet{wright10}.} and $\langle$$R_{H2}$$\rangle$.  With the relatively low Spearman correlation coefficient ($\rho$~= 0.326) and high probability of non-correlation (n = 0.079), we do not find a correlation between the slope of the inner disk dust SED and the average H$_{2}$ emitting radius.   This is in some tension with the findings of~\citet{hoadley15}, who observed a positive correlation between their modeled radial distribution of H$_{2}$ fluorescence and the $n_{13-31}$ dust SED derived from $Spitzer$-IRS spectroscopy.   There are two key differences between these studies: first, the $Spitzer$ data employed by ~\citet{hoadley15} are a cleaner metric of warm inner disk dust clearing than the $WISE$ photometric points~\citep{espaillat14}.  The $WISE$ W3 band in particular includes the 10~$\mu$m silicate emission feature seen in many CTTS spectra~\citep{olofsson10}, and the lower angular resolution of $WISE$ increases the possibility of source confusion.  Second, \citet{hoadley15} employ an emission line forward modeling analysis of a higher-S/N sample of disk targets; they find correlations between inner disk dust clearing and the maxima of the H$_{2}$ radial distributions derived from their models.   A follow-on study of the high-S/N ULLYSES sources with $Spitzer$ or $JWST$ spectra would be valuable for a more direct comparison with the~\citet{hoadley15} results.

\subsection{Inner Disk Opacity and H$_{2}$ Continuum Emission}

There are two proposed origins for the 1600~\AA\ Bump emission seen in many protoplanetary disk sources. The first proposes that free electrons (possibly liberated by X-ray emission from the central star) collide with the disk surface population of H$_{2}$, creating an electron impact spectrum similar to that observed in the aurorae of giant planets~\citep{bergin04,ingleby09}.  The second route for the formation of the 1600~\AA\ Bump proposes that bright stellar+accretion-generated Ly$\alpha$ emission dissociates water molecules and that the broad continuum Bump is produced during the photodissociation of the non-thermal H$_{2}$ created during the break-up (dissociation) of the water molecules~\citep{harrevelt08, france17}.  

Analysis of the ULLYSES DR5 sample finds 24 sources with detectable 1600~\AA\ Bump emission, a detection rate of 38\%, roughly consistent with the $\approx$~50\% detection rate reported in \citet{france17}.  We find a strong correlation between the strength of the Bump and the Ly$\alpha$ field strength.  Figure 10 illustrates the relationship between the Ly$\alpha$ and 1600~\AA\ Bump luminosity, the highly significant correlation ($\rho$~=~0.8, n~=~2.7~$\times$~10$^{-6}$) strongly suggests Ly$\alpha$ photons as central to the formation of the H$_{2}$ continuum emission feature.  For comparison, the correlation between \ion{C}{4} and the 1600~\AA\ Bump luminosity is less pronounced ($\rho$~=~0.5, n~=~1.6~$\times$~10$^{-2}$).  We also note that the Ly$\alpha$ luminosities in sources with strong Bump detections are $\gtrsim$ 5~$\times$~10$^{31}$ erg s$^{-1}$, the Ly$\alpha$ region strongly anchored by the~\citet{france12} Ly$\alpha$ reconstruction sample.   

\begin{figure}[htbp]
   \centering
   \includegraphics[scale=.48,clip,trim=0mm 0mm 5mm 5mm,angle=0]{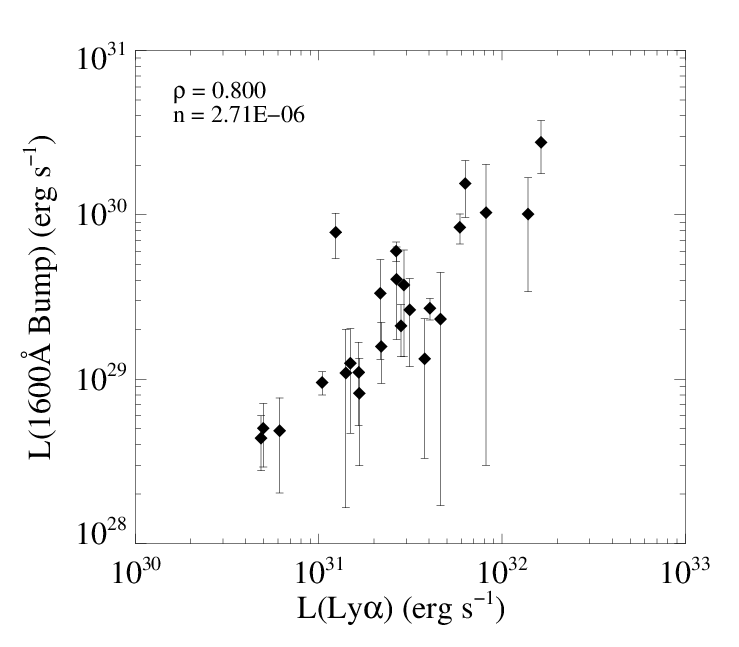}
   \figcaption{The 1600\AA\ Bump luminosity is strongly correlated with Ly$\alpha$ luminosity. This correlation supports the hypothesis that the Bump is powered by Ly$\alpha$-driven H$_{2}$O dissociation. The Spearman rank coefficient ($\rho$) and probability coefficient for null correlation (n) are shown at the upper left.  }
\end{figure}

\begin{figure}[htbp]
   \centering
   \includegraphics[scale=.48,clip,trim=0mm 0mm 5mm 5mm,angle=0]{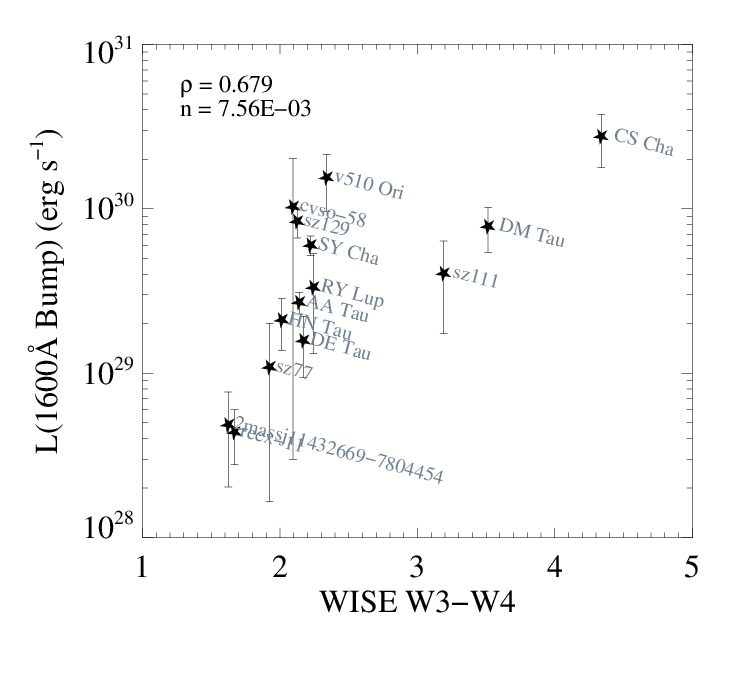}
   \figcaption{A comparison between the $WISE$ 12$\mu$m~-~22$\mu$m color and the 1600\AA\ Bump luminosity.  Targets are labeled for reference and the Spearman rank coefficient ($\rho$) and probability coefficient for null correlation (n) are shown at the upper left.  The positive correlation indicates that as the inner disk opacity decreases, stellar+accretion-generated Ly$\alpha$ photons can penetrate farther into the disk where they dissociate H$_{2}$O, creating the non-thermal H$_{2}$ population that gives rise to the 1600~\AA\ molecular dissociation continuum.   }
\end{figure}

Within the sample of 24 1600~\AA\ Bump detections, 14 had $WISE$ W3-W4 colors without data quality warnings.   We plot the Bump luminosity versus the W3-W4 color in Figure 11.  We observe a correlation ($\rho$~=~0.68, n~=~7.6~$\times$~10$^{-3}$) between these quantities, with disks showing the largest depletion of optically thick warm dust (noting the caveats about spectral and spatial contamination in the $WISE$ photometric bands discussed previously) having the largest H$_{2}$ continuum emission luminosity.   Taken together with the strong correlation with L(Ly$\alpha$), the emerging picture for the formation of the 1600~\AA\ Bump includes both strong accretion-generated Ly$\alpha$ flux and low optical depth in the inner disk gas and dust environment.  Bright Ly$\alpha$ emission alone is not a sufficient condition to produce the H$_{2}$ continuum emission, as there are 44 targets in our sample with L(Ly$\alpha$)~$>$~10$^{31}$ erg s$^{-1}$ (including, e.g., CVSO 109, Figure 3 and \citealt{espaillat22}) yet only 21 of those are detected in the 1600~\AA\ Bump feature.  The second required condition is that those Ly$\alpha$ photons are able to reach the molecular reservoirs in the inner disk.   As the gas and dust opacities decline with time, Ly$\alpha$ photons can more readily reach the molecular populations of the inner disk, where they dissociate water molecules and the resultant non-thermal H$_{2}$ gas.  

\section{Summary}

We have analyzed far-ultraviolet spectra of 71 young stars included in the ULLYSES Data Release 5.   We have created a uniform emission line extraction and measurement database of fluorescent H$_{2}$ and magnetospheric and accretion-generated hot gas lines that are hosted on a public website at the University of Colorado$^{3}$.
The H$_{2}$  database is used to compute the total fluorescent H$_{2}$ luminosity and to estimate the Ly$\alpha$ luminosity from the ULLYSES sample.   

The line-widths are combined with the stellar mass and disk inclination to derive the average H$_{2}$ emitting radius for the fluorescent emission.  We find that the majority of the fluorescent emission arises between 0.1 and 1.4 au from the central star.  We find a weak positive correlation between the average emitting radius and the stellar FUV flux, and a stronger correlation between the H$_{2}$ emitting radius and the dust disk mass.  We extracted FUV continuum spectra and H$_{2}$ dissociation emission from the ULLYSES sample, finding 24 detections of H$_{2}$ continuum emission.  We find strong correlations between the H$_{2}$ continuum emission and both the Ly$\alpha$ luminosity and the mid-IR (10~--~22~$\mu$m) inner disk dust slope.  These results support a picture for the production of the H$_{2}$ continuum where a decrease of the inner disk gas and dust opacity enables bright stellar Ly$\alpha$ emission to reach the molecular populations of the inner disk \citep{bergin03,bethell11}, driving molecular dissociation and the associated continuum emission.

\acknowledgments
  This work was carried out as part of the Outflows and Disks around Young Stars: Synergies for the Exploration of Ullyses Spectra (ODYSSEUS) program, supported by $HST$ archival grant number 16129-025 to the University of Colorado at Boulder.  E.F. acknowledges support from the project PRIN‐INAF 2019 “Spectroscopically Tracing the Disk Dispersal Evolution (STRADE)”.  We acknowledge the excellent work of the ULLYSES implementation team at STScI to develop and execute this ambitious spectroscopic observing program.   The HST-ULLYSES data presented in this paper were obtained from the Mikulski Archive for Space Telescopes (MAST) at the Space Telescope Science Institute. The specific observations analyzed can be accessed via \dataset[DOI: 10.17909/t9-jzeh-xy14]{https://doi.org/10.17909/t9-jzeh-xy14}.

\bibliography{H2ullys_bibtex}{}
\bibliographystyle{aasjournal}

\startlongtable
\begin{deluxetable*}{lcccccccc}
\tabletypesize{\scriptsize}
\tablewidth{0pt}
\tablecaption{ULLYSES DR5 Target Parameters}
\tablehead{
\colhead{Name} & \colhead{M$_{*}$} & \colhead{$\dot M_{acc}$} & \colhead{$i$} & \colhead{$d$}  & \colhead{A$_{V}$}  &  \colhead{$WISE$ W3}   &  \colhead{$WISE$ W4} &  \colhead{References} \\
\colhead{} & \colhead{(M$_{\odot}$)} & \colhead{(10$^{-8}$ M$_{\odot}$ yr$^{-1}$)} & \colhead{($deg$)} & \colhead{(pc)}  & \colhead{(mag)}  &  \colhead{(mag)}   &  \colhead{(mag)}  &   \colhead{ }
}
\startdata
2MASS-j04390163+2336029  &  0.17  &  0.04  &  $\cdots$  &  126  &  0.00  &  $\cdots$  &  $\cdots$   & 14,15   \\
2MASSj11432669-7804454  &  0.14  &  0.19  &  $\cdots$  &  153  &  0.40  &  8.676  &  7.052   &  1    \\
CHX18n  &  0.81  &  0.47  &  56  &  192  &  0.80  &  5.560  &  3.584   &    1,29  \\
CVSO-17  &  0.37  &  0.00  &  $\cdots$  &  414  &  0.00  &  11.130  &  8.932   &   7    \\
CVSO-36  &  0.39  &  0.00  &  $\cdots$  &  335  &  0.10  &  10.306  &  8.593   &   7   \\
CVSO-58  &  0.81  &  0.43  &  $\cdots$  &  349  &  0.80  &  7.000  &  4.905   &   7   \\
CVSO-90  &  0.62  &  0.25  &  $\cdots$  &  338  &  0.10  &  7.622  &  5.345   &   7   \\
CVSO-104  &  0.37  &  0.32  &  43  &  360  &  0.20  &  7.560  &  5.571   &   7,20    \\
CVSO-107  &  0.53  &  5.01  &  $\cdots$  &  330  &  0.30  &  7.419  &  5.128   &   7   \\
CVSO-109  &  0.46  &  3.24  &  $\cdots$  &  400  &  0.10  &  6.523  &  4.527   &   7   \\
CVSO-146  &  0.86  &  0.27  &  $\cdots$  &  332  &  0.60  &  7.072  &  4.597   &  7    \\
CVSO-165  &  0.84  &  0.08  &  $\cdots$  &  400  &  0.20  &  6.571  &  4.626   &   7   \\
CVSO-176  &  0.25  &  1.45  &  $\cdots$  &  302  &  1.00  &  7.896  &  6.221   &   7   \\
echa-j0843-7915 (RECX-15)  &  0.20  &  0.08  &  60  &  103  &  0.00  &  $\cdots$  &  $\cdots$   &   4,5,27     \\
echa-j0844-7833  &  0.07  &  0.00  &  $\cdots$  &  98  &  0.50  &  $\cdots$  &  $\cdots$   &   4,5    \\
LkCa-4  &  0.77  &  0.00  &  $\cdots$  &  129  &  0.69  &  8.038  &  7.934   &   9,10   \\
LkCa-15  &  0.85  &  0.13  &  49  &  157  &  0.60  &  5.696  &  3.565   &    11,3,19   \\
LkCa-19  &  1.35  &  0.00  &  $\cdots$  &  157  &  0.00  &  7.972  &  7.109   &   9   \\
HN5  &  0.18  &  0.28  &  $\cdots$  &  195  &  1.10  &  5.864  &  4.371   &   1   \\
RECX-1  &  0.75  &  0.00  &  $\cdots$  &  98  &  0.00  &  7.058  &  6.943    &  4,5   \\
RECX-11  &  0.83  &  0.02  &  70  &  99  &  0.10  &  5.388  &  3.718    &   4,5,27   \\
RXJ0438.6+1556  &  1.20  &  0.03  &  $\cdots$  &  139  &  0.30  &  8.154  &  8.122   &   13  \\
RXJ1556.1-3655  &  0.50  &  1.20  &  53  &  158  &  1.00  &  6.268  &  3.887   &  16,30    \\
SSTC2dj160000-422158  &  0.19  &  0.06  &  65  &  159  &  0.10  &  $\cdots$  &  $\cdots$   &   16,29   \\
SSTC2dj160830-382827  &  1.40  &  0.11  &  72  &  153  &  0.20  &  $\cdots$  &  $\cdots$    &   16, 17   \\
SSTC2dj161243-381503  &  0.44  &  0.17  &  43  &  159  &  0.80  &  $\cdots$  &  $\cdots$   &    16,30  \\
SSTC2dj161344-373646  &  0.16  &  0.11  &  54  &  158  &  0.60  &  $\cdots$  &  $\cdots$    &   16,29  \\
Sz 19  &  2.08  &  2.34  &  45  &  189  &  1.50  &  3.011  &  1.057   &    1,29  \\
Sz 45  &  0.56  &  0.51  &  43  &  189  &  0.70  &  6.948  &  4.156   &   1,29   \\
Sz 66  &  0.29  &  0.05  &  40  &  156  &  1.00  &  6.363  &  4.432   &  6,29    \\
Sz 68  &  1.40  &  0.58  &  32  &  152  &  1.00  &  2.896  &  0.856   &    16,30   \\
Sz 69  &  0.20  &  0.03  &  35  &  155  &  0.00  &  5.614  &  4.420   &   6,28    \\
Sz 71  &  0.37  &  0.11  &  40  &  156  &  0.70  &  5.716  &  3.671   &    6,28   \\
Sz 72  &  0.37  &  0.95  &  53  &  156  &  1.00  &  5.987  &  3.911   &   6,28    \\
Sz 75  &  0.82  &  6.79  &  $\cdots$  &  152  &  1.00  &  4.342  &  2.379   &  16   \\
Sz 76  &  0.22  &  0.07  &  $\cdots$  &  159  &  0.30  &  7.366  &  5.274   &   16   \\
Sz 77  &  0.75  &  0.07  &  $\cdots$  &  155  &  0.30  &  5.362  &  3.437   &   16   \\
Sz 84  &  0.16  &  0.06  &  75  &  155  &  0.00  &  9.115  &  6.340   &    6,29  \\
Sz 97  &  0.23  &  0.03  &  73  &  157  &  0.00  &  7.749  &  4.976   &    6,29  \\
Sz 98  &  0.70  &  5.89  &  46  &  156  &  1.00  &  3.986  &  2.097   &   16,29   \\
Sz 99  &  0.23  &  0.04  &  $\cdots$  &  158  &  0.00  &  7.847  &  6.009   &   6  \\
Sz 100  &  0.16  &  0.04  &  47  &  141  &  0.00  &  6.486  &  4.517   &    6,29  \\
Sz 102  &  0.24  &  0.08  &  53  &  160  &  1.13  &  6.196  &  3.382   &   16,29   \\
Sz 103  &  0.22  &  0.10  &  50  &  157  &  0.70  &  6.940  &  4.843   &    6,29  \\
Sz 104  &  0.16  &  0.02  &  57  &  159  &  0.00  &  7.438  &  4.806   &    6,29  \\
Sz 110  &  0.22  &  0.30  &  32  &  157  &  0.00  &  6.307  &  4.116   &   6,29   \\
Sz 111  &  0.58  &  0.09  &  54  &  158  &  0.50  &  8.922  &  5.731   &   6,29   \\
Sz 114  &  0.21  &  0.11  &  6  &  153  &  0.30  &  5.721  &  3.380    &  6,29   \\
Sz 117  &  0.26  &  0.25  &  47  &  156  &  0.50  &  7.169  &  4.884    &   16,29  \\
Sz 129  &  0.79  &  0.40  &  31  &  160  &  0.90  &  5.722  &  3.597   &  16,29   \\
Sz 130  &  0.36  &  0.15  &  36  &  159  &  0.40  &  6.530  &  4.514    &  6,28    \\
AA Tau  &  0.80  &  0.33  &  59  &  134  &  0.50  &  4.645  &  2.505    &  11,3,18    \\
CE Ant  &  0.46  &  0.00  &  $\cdots$  &  34  &  0.00  &  6.598  &  5.998    &  3   \\
CS Cha  &  1.05  &  1.20  &  37  &  168  &  0.80  &  7.095  &  2.758    &   2,3,13,19   \\
DE Tau  &  0.59  &  2.64  &  34  &  128  &  0.60  &  4.895  &  2.725    &   11,3,21   \\
DK Tau  &  0.71  &  3.79  &  58  &  140  &  0.80  &  3.599  &  1.800    &  11,3,22    \\
DM Tau  &  0.50  &  0.29  &  35  &  144  &  0.00  &  7.085  &  3.571    &   2,3,24   \\
DN Tau  &  0.60  &  0.35  &  35  &  128  &  1.90  &  5.166  &  3.039     &  11,3,23   \\
DR Tau  &  0.80  &  3.16  &  5  &  193  &  0.48  &  3.008  &  1.066    &   11,3,30   \\
HN Tau  &  0.85  &  0.13  &  50  &  140  &  0.50  &  4.174  &  2.160    &  11,3,31   \\
IN Cha  &  0.19  &  0.05  &  $\cdots$  &  193  &  0.20  &  7.218  &  5.316    &   1  \\
IP Tau  &  0.68  &  0.08  &  35  &  129  &  0.20  &  5.436  &  3.578    &   11,3,31  \\
MY Lup  &  1.06  &  0.02  &  72  &  157  &  1.30  &  5.164  &  2.833    &   16,26   \\
RY Lup  &  1.71  &  1.00  &  68  &  153  &  0.40  &  3.647  &  1.404   &   16,19    \\
SY Cha  &  0.78  &  0.04  &  51  &  180  &  0.50  &  5.435  &  3.212    &   1,29  \\
TX Ori  &  1.09  &  6.17  &  $\cdots$  &  385  &  0.40  &  5.101  &  3.083    &   8  \\
UX Tau A  &  1.30  &  1.00  &  35  &  140  &  0.20  &  5.710  &  2.176   &  12,19     \\
V397 Aur  &  0.70  &  0.00  &  $\cdots$  &  165  &  0.20  &  7.854  &  6.743   &  9    \\
XX Cha  &  0.29  &  0.10  &  $\cdots$  &  192  &  0.30  &  5.751  &  3.569    &  1   \\
V505 Ori  &  0.81  &  0.30  &  $\cdots$  &  392  &  1.00  &  6.564  &  4.539    &  8   \\
V510 Ori  &  0.76  &  0.58  &  $\cdots$  &  390  &  0.10  &  5.598  &  3.258   &   8 
\enddata
\tablecomments{References for the stellar masses, reddening, accretion rates, and disk inclination:
(1) \citep{manara17}, 
(2) \citep{furlan09},  
(3) \citep{ingleby13}, 
(4) \citep{rugel18},   
(5) \citep{Luhman2004}, 
(6) \citep{alcala14}, 
(7) \citep{calvet05}, 
(8) \citep{mauco16}, 
(9) \citep{kenyon95}, 
(10) \citep{donati14}
(11) \citep{furlan11}
(12) \citep{costigan14}, 
(13) \citep{manara17b}, 
(14) \citep{slesnick06}, 
(15) \citep{Herczeg2008}, 
(16) \citep{alcala17}, 
(17) \citep{vandermarel18},
(18) \citep{loomis17}, 
(19) \citep{Francis2020},
(20) \citep{Frasca2021},
(21) \citep{Simon2019}, 
(22) \citep{nelissen2023},
(23) \citep{Long2019},  
(24) \citep{Kudo2018}, 
(25) \citep{Kurtovic2018}, 
(26) \citep{Huang2018}, 
(27) \citep{Lawson2004},
(28) \citep{Ansdell2016},
(29) \citep{Hendler20},
(30) \citep{Braun21},
(31) \citep{simon17}
}
\end{deluxetable*}

\clearpage

\startlongtable
\begin{longrotatetable}
\begin{deluxetable}{lccccccccc}
\tabletypesize{\scriptsize}
\tablewidth{0pt}
\tablecaption{ULLYSES FUV Measurements}
\tablehead{
\colhead{Name} & \colhead{L(H$_{2}$)} & \colhead{L(Ly$\alpha$)} & \colhead{L(\ion{C}{4})} & \colhead{L(1600\AA\ Bump)}  & \colhead{L(FUV Cont)} & \colhead{FWHM(H$_{2}$)$_{[1,4]}$}  &  \colhead{FWHM(H$_{2}$)$_{[0,1]}$}   &  \colhead{$<R(H_{2})>_{[1,4]}$}   &
\colhead{$<R(H_{2})>_{[0,1]}$}  \\
\colhead{} & \colhead{(10$^{29}$ erg s$^{-1}$)} & \colhead{(10$^{29}$ erg s$^{-1}$)} & \colhead{(10$^{29}$ erg s$^{-1}$)} & \colhead{(10$^{29}$ erg s$^{-1}$)} & \colhead{(10$^{29}$ erg s$^{-1}$)}   & \colhead{(km s$^{-1}$)}  &  \colhead{(km s$^{-1}$)}   &  \colhead{(au)} &  \colhead{(au)}
}
\startdata
2MASS-j04390163+2336029  &  0.05 $\pm$ 0.01  &  2.89  &  0.02  &  $\cdots$ $\pm$ $\cdots$  &  0.1 & 76.4 $\pm$ 17.3  &  13.4 $\pm$ 29.0  &  0.08 $\pm$ 0.02  &  2.52 $\pm$ 5.46  \\
2MASSj11432669-7804454  &  2.20 $\pm$ 0.22  &  61.18  &  0.65  &  0.49 $\pm$ 0.28  &  3.6  & 61.7 $\pm$ 14.8  &  67.0 $\pm$ 4.5  &  0.10 $\pm$ 0.02  &  0.08 $\pm$ 0.01  \\
CHX18n  &  17.65 $\pm$ 2.75  &  348.59  &  8.38  &  $\cdots$ $\pm$ $\cdots$  &  36.8 &  51.8 $\pm$ 4.9  &  49.1 $\pm$ 4.5  &  0.75 $\pm$ 0.07  &  0.83 $\pm$ 0.08  \\
CVSO-17  &  0.04 $\pm$ 0.07  &  15.01  &  0.17  &  $\cdots$ $\pm$ $\cdots$  &  4.2 & $\cdots$ $\pm$ $\cdots$  &  $\cdots$ $\pm$ $\cdots$  &  $\cdots$ $\pm$ $\cdots$  &  $\cdots$ $\pm$ $\cdots$  \\
CVSO-36  &  0.11 $\pm$ 0.09  &  11.62  &  0.09  &  $\cdots$ $\pm$ $\cdots$  &  2.8 &  $\cdots$ $\pm$ $\cdots$  &  $\cdots$ $\pm$ $\cdots$  &  $\cdots$ $\pm$ $\cdots$  &  $\cdots$ $\pm$ $\cdots$  \\
CVSO-58  &  37.82 $\pm$ 3.82  &  817.71  &  21.69  &  10.30 $\pm$ 10.00  &  277.6   & 54.3 $\pm$ 2.5  &  44.7 $\pm$ 4.4  &  0.73 $\pm$ 0.03  &  1.08 $\pm$ 0.11  \\
CVSO-90  &  16.74 $\pm$ 1.74  &  423.96  &  10.30  &  $\cdots$ $\pm$ $\cdots$  &    79.1   & 51.8 $\pm$ 4.9  &  51.7 $\pm$ 1.6  &  0.62 $\pm$ 0.06  &  0.62 $\pm$ 0.02  \\
CVSO-104  &  8.66 $\pm$ 0.69  &  258.39  &  4.14  &  $\cdots$ $\pm$ $\cdots$  &  51.8  &   49.3 $\pm$ 2.5  &  53.7 $\pm$ 4.2  &  0.25 $\pm$ 0.01  &  0.21 $\pm$ 0.02  \\
CVSO-107  &  15.21 $\pm$ 2.49  &  388.81  &  7.94  &  $\cdots$ $\pm$ $\cdots$  &  136.0  &  64.1 $\pm$ 7.4  &  53.6 $\pm$ 13.4  &  0.34 $\pm$ 0.04  &  0.49 $\pm$ 0.12  \\
CVSO-109  &  14.60 $\pm$ 1.52  &  407.72  &  12.72  &  $\cdots$ $\pm$ $\cdots$  &  229.5  &  44.4 $\pm$ 2.5  &  40.7 $\pm$ 3.0  &  0.62 $\pm$ 0.03  &  0.74 $\pm$ 0.05  \\
CVSO-146  &  28.78 $\pm$ 4.50  &  645.24  &  15.46  &  $\cdots$ $\pm$ $\cdots$  &  310.0  &  51.8 $\pm$ 19.7  &  53.6 $\pm$ 6.7  &  0.85 $\pm$ 0.32  &  0.80 $\pm$ 0.10  \\
CVSO-165  &  11.39 $\pm$ 2.00  &  335.04  &  5.63  &  $\cdots$ $\pm$ $\cdots$  &  66.2   &   76.4 $\pm$ 2.5  &  53.6 $\pm$ 13.4  &  0.38 $\pm$ 0.01  &  0.78 $\pm$ 0.19  \\
CVSO-176  &  51.99 $\pm$ 5.58  &  990.56  &  10.54  &  $\cdots$ $\pm$ $\cdots$  &  245.9   &    74.0 $\pm$ 2.5  &  46.9 $\pm$ 2.2  &  0.12 $\pm$ 0.00  &  0.30 $\pm$ 0.01  \\
echa-j0843-7915  &  5.33 $\pm$ 0.48  &  104.36  &  0.48  &  0.96 $\pm$ 0.15  &  5.5  &   49.3 $\pm$ 2.5  &  55.8 $\pm$ 2.2  &  0.22 $\pm$ 0.01  &  0.17 $\pm$ 0.01  \\
echa-j0844-7833'  &  0.15 $\pm$ 0.06  &  6.21  &  0.21  &  $\cdots$ $\pm$ $\cdots$  &   0.7   &    $\cdots$ $\pm$ $\cdots$  &  $\cdots$ $\pm$ $\cdots$  &  $\cdots$ $\pm$ $\cdots$  &  $\cdots$ $\pm$ $\cdots$  \\
LkCa-4  &  0.07 $\pm$ 0.30  &  3.45  &  1.50  &  $\cdots$ $\pm$ $\cdots$  &     2.3   &     $\cdots$ $\pm$ $\cdots$  &  $\cdots$ $\pm$ $\cdots$  &  $\cdots$ $\pm$ $\cdots$  &  $\cdots$ $\pm$ $\cdots$  \\
LkCa-15  &  23.78 $\pm$ 3.11  &  405.89  &  12.40  &  $\cdots$ $\pm$ $\cdots$  &  50.5    &         51.8 $\pm$ 2.5  &  53.6 $\pm$ 2.2  &  0.64 $\pm$ 0.03  &  0.60 $\pm$ 0.02  \\
LkCa-19  &  0.08 $\pm$ 0.08  &  4.31  &  0.71  &  $\cdots$ $\pm$ $\cdots$  &  1.8   &     $\cdots$ $\pm$ $\cdots$  &  $\cdots$ $\pm$ $\cdots$  &  $\cdots$ $\pm$ $\cdots$  &  $\cdots$ $\pm$ $\cdots$ \\
HN5  &  2.52 $\pm$ 0.35  &  75.19  &  0.41  &  $\cdots$ $\pm$ $\cdots$  &    5.4   &    133.2 $\pm$ 93.7  &  17.9 $\pm$ 2.2  &  0.03 $\pm$ 0.02  &  1.50 $\pm$ 0.19  \\
RECX-1  &  -0.10 $\pm$ 0.13  &  0.07  &  0.51  &  $\cdots$ $\pm$ $\cdots$  &    2.3    &     $\cdots$ $\pm$ $\cdots$  &  $\cdots$ $\pm$ $\cdots$  &  $\cdots$ $\pm$ $\cdots$  &  $\cdots$ $\pm$ $\cdots$  \\
RECX-11  &  2.07 $\pm$ 0.26  &  48.42  &  2.35  &  0.44 $\pm$ 0.16  &    5.1   &    61.6 $\pm$ 2.5  &  65.9 $\pm$ 2.0  &  0.69 $\pm$ 0.03  &  0.60 $\pm$ 0.02  \\
RXJ0438.6+1556  &  0.06 $\pm$ 0.10  &  3.10  &  1.01  &  $\cdots$ $\pm$ $\cdots$  &    2.7   &     182.5 $\pm$ 88.8  &  78.2 $\pm$ 67.0  &  0.10 $\pm$ 0.05  &  0.52 $\pm$ 0.45  \\
RXJ1556.1-3655  &  105.72 $\pm$ 9.49  &  1323.61  &  53.50  &  $\cdots$ $\pm$ $\cdots$  &  116.5   &     56.6 $\pm$ 2.5  &  49.1 $\pm$ 2.2  &  0.36 $\pm$ 0.02  &  0.48 $\pm$ 0.02  \\
SSTC2dj160000-422158  &  0.14 $\pm$ 0.01  &  6.88  &  0.02  &  $\cdots$ $\pm$ $\cdots$  &   0.5  &   61.6 $\pm$ 2.5  &  35.7 $\pm$ 20.1  &  0.15 $\pm$ 0.01  &  0.44 $\pm$ 0.25  \\
SSTC2dj160830-382827  &  4.83 $\pm$ 0.75  &  113.89  &  1.28  &  $\cdots$ $\pm$ $\cdots$  &    6.6   &   46.8 $\pm$ 2.5  &  40.2 $\pm$ 20.1  &  2.06 $\pm$ 0.11  &  2.79 $\pm$ 1.39  \\
SSTC2dj161243-381503  &  7.63 $\pm$ 0.46  &  166.38  &  3.83  &  0.82 $\pm$ 0.52  &    14.0   &    54.3 $\pm$ 4.9  &  42.4 $\pm$ 2.2  &  0.25 $\pm$ 0.02  &  0.41 $\pm$ 0.02  \\
SSTC2dj161344-373646  &  2.17 $\pm$ 0.51  &  61.26  &  4.84  &  $\cdots$ $\pm$ $\cdots$  &  13.2   &      49.3 $\pm$ 2.5  &  35.7 $\pm$ 13.4  &  0.16 $\pm$ 0.01  &  0.30 $\pm$ 0.11  \\
Sz 19  &  8.29 $\pm$ 1.69  &  190.54  &  13.98  &  $\cdots$ $\pm$ $\cdots$  &  42.1   &    49.3 $\pm$ 9.9  &  41.3 $\pm$ 4.5  &  1.44 $\pm$ 0.29  &  2.05 $\pm$ 0.22  \\
Sz 45  &  21.74 $\pm$ 1.96  &  408.30  &  8.43  &  $\cdots$ $\pm$ $\cdots$  &  66.8   &    46.8 $\pm$ 2.5  &  42.4 $\pm$ 2.2  &  0.43 $\pm$ 0.02  &  0.53 $\pm$ 0.03  \\
Sz 66  &  17.24 $\pm$ 1.59  &  313.71  &  0.11  &  2.64 $\pm$ 1.45  &  11.9   &    44.4 $\pm$ 7.4  &  44.7 $\pm$ 2.2  &  0.22 $\pm$ 0.04  &  0.22 $\pm$ 0.01  \\
Sz 68  &  15.88 $\pm$ 2.39  &  291.37  &  8.66  &  3.74 $\pm$ 2.37  &   30.3   &   59.2 $\pm$ 12.3  &  58.1 $\pm$ 6.7  &  0.42 $\pm$ 0.09  &  0.43 $\pm$ 0.05  \\
Sz 69  &  1.69 $\pm$ 0.21  &  49.80  &  0.03  &  0.50 $\pm$ 0.21  &    1.1   &     54.3 $\pm$ 4.9  &  42.4 $\pm$ 2.2  &  0.08 $\pm$ 0.01  &  0.13 $\pm$ 0.01  \\
Sz 71  &  6.71 $\pm$ 0.74  &  148.76  &  2.58  &  1.25 $\pm$ 0.78  &    10.8  &    46.9 $\pm$ 2.5  &  38.0 $\pm$ 4.4  &  0.25 $\pm$ 0.01  &  0.38 $\pm$ 0.04  \\
Sz 72  &  11.27 $\pm$ 1.98  &  224.08  &  14.26  &  $\cdots$ $\pm$ $\cdots$  &   65.0  &   56.7 $\pm$ 9.9  &  45.4 $\pm$ 1.9  &  0.26 $\pm$ 0.05  &  0.41 $\pm$ 0.02  \\
Sz 75  &  162.14 $\pm$ 25.70  &  1826.48  &  73.94  &  $\cdots$ $\pm$ $\cdots$  &  966.3  &    78.9 $\pm$ 4.9  &  60.3 $\pm$ 6.7  &  0.35 $\pm$ 0.02  &  0.61 $\pm$ 0.07  \\
Sz 76  &  1.33 $\pm$ 0.14  &  41.62  &  0.27  &  $\cdots$ $\pm$ $\cdots$  &   2.3   &      37.0 $\pm$ 2.5  &  60.3 $\pm$ 13.4  &  0.43 $\pm$ 0.03  &  0.16 $\pm$ 0.04  \\
Sz 77  &  6.27 $\pm$ 0.78  &  140.53  &  2.77  &  1.09 $\pm$ 0.92  &   16.7  &     54.3 $\pm$ 9.9  &  46.9 $\pm$ 64.8  &  0.68 $\pm$ 0.12  &  0.91 $\pm$ 1.25  \\
Sz 84  &  0.62 $\pm$ 0.09  &  22.58  &  0.10  &  $\cdots$ $\pm$ $\cdots$  &  0.6   &    44.4 $\pm$ 4.9  &  46.9 $\pm$ 11.1  &  0.27 $\pm$ 0.03  &  0.24 $\pm$ 0.06  \\
Sz 97  &  0.29 $\pm$ 0.05  &  12.25  &  0.26  &  $\cdots$ $\pm$ $\cdots$  &  1.6  &    51.8 $\pm$ 2.5  &  105.0 $\pm$ 22.4  &  0.28 $\pm$ 0.01  &  0.07 $\pm$ 0.01  \\
Sz 98  &  28.07 $\pm$ 2.34  &  461.61  &  4.79  &  2.32 $\pm$ 2.15  &    59.7   &    49.3 $\pm$ 2.5  &  44.7 $\pm$ 2.2  &  0.54 $\pm$ 0.03  &  0.66 $\pm$ 0.03  \\
Sz 99  &  0.25 $\pm$ 0.05  &  10.95  &  0.41  &  $\cdots$ $\pm$ $\cdots$  &  1.7    &  61.5 $\pm$ 17.2  &  22.3 $\pm$ 58.0  &  0.16 $\pm$ 0.05  &  1.23 $\pm$ 3.19  \\
Sz 100  &  1.30 $\pm$ 0.12  &  39.04  &  0.17  &  $\cdots$ $\pm$ $\cdots$  &    1.6    &     46.8 $\pm$ 2.5  &  15.8 $\pm$ 2.1  &  0.14 $\pm$ 0.01  &  1.22 $\pm$ 0.17  \\
Sz 102  &  111.34 $\pm$ 12.58  &  1386.21  &  44.74  &  10.10 $\pm$ 6.69  &  379.8   &    59.1 $\pm$ 4.9  &  62.5 $\pm$ 2.2  &  0.16 $\pm$ 0.01  &  0.14 $\pm$ 0.00  \\
Sz 103  &  7.63 $\pm$ 0.66  &  165.16  &  0.22  &  $\cdots$ $\pm$ $\cdots$  &   27.0  &   49.3 $\pm$ 2.5  &  53.6 $\pm$ 2.2  &  0.19 $\pm$ 0.01  &  0.16 $\pm$ 0.01  \\
Sz 104  &  0.14 $\pm$ 0.07  &  6.95  &  0.21  &  $\cdots$ $\pm$ $\cdots$  &    3.5  &     76.3 $\pm$ 27.1  &  136.2 $\pm$ 109.4  &  0.07 $\pm$ 0.02  &  0.02 $\pm$ 0.02  \\
Sz 110  &  2.00 $\pm$ 0.35  &  57.43  &  0.72  &  $\cdots$ $\pm$ $\cdots$  &  2.5  &       68.9 $\pm$ 9.9  &  100.5 $\pm$ 29.0  &  0.05 $\pm$ 0.01  &  0.02 $\pm$ 0.01  \\
Sz 111  &  13.88 $\pm$ 1.54  &  265.72  &  3.30  &  4.05 $\pm$ 2.31  &   16.6   &   46.8 $\pm$ 2.5  &  58.1 $\pm$ 6.7  &  0.63 $\pm$ 0.03  &  0.41 $\pm$ 0.05  \\
Sz 114  &  12.68 $\pm$ 0.97  &  244.59  &  1.51  &  $\cdots$ $\pm$ $\cdots$  &   5.6  &       44.3 $\pm$ 2.5  &  35.7 $\pm$ 2.2  &  0.01 $\pm$ 0.00  &  0.01 $\pm$ 0.00  \\
Sz 117  &  1.36 $\pm$ 0.82  &  42.13  &  2.54  &  $\cdots$ $\pm$ $\cdots$  &   8.1  &       73.8 $\pm$ 152.6  &  78.1 $\pm$ 73.7  &  0.09 $\pm$ 0.19  &  0.08 $\pm$ 0.08  \\
Sz 129  &  37.69 $\pm$ 4.56  &  588.70  &  18.31  &  8.39 $\pm$ 1.74  &  190.6   &    51.8 $\pm$ 2.5  &  49.1 $\pm$ 2.2  &  0.28 $\pm$ 0.01  &  0.31 $\pm$ 0.01  \\
Sz 130  &  2.28 $\pm$ 0.35  &  63.87  &  1.33  &  $\cdots$ $\pm$ $\cdots$  &    5.0  &  49.3 $\pm$ 2.5  &  42.4 $\pm$ 22.3  &  0.19 $\pm$ 0.01  &  0.25 $\pm$ 0.13  \\
AA Tau  &  25.65 $\pm$ 2.60  &  403.91  &  2.04  &  2.70 $\pm$ 0.41  &  16.6   &     54.2 $\pm$ 2.5  &  53.6 $\pm$ 2.2  &  0.71 $\pm$ 0.03  &  0.73 $\pm$ 0.03  \\
CE Ant  &  0.00 $\pm$ 0.01  &  0.13  &  0.06  &  $\cdots$ $\pm$ $\cdots$  &   0.3   &     $\cdots$ $\pm$ $\cdots$  &  $\cdots$ $\pm$ $\cdots$  &  $\cdots$ $\pm$ $\cdots$  &  $\cdots$ $\pm$ $\cdots$  \\
CS Cha  &  133.10 $\pm$ 12.53  &  1632.53  &  0.44  &  27.60 $\pm$ 9.81  &  330.3    &    35.6 $\pm$ 0.3  &  35.7 $\pm$ 2.2  &  1.06 $\pm$ 0.01  &  1.06 $\pm$ 0.07  \\
DE Tau  &  12.17 $\pm$ 2.51  &  219.19  &  6.96  &  1.58 $\pm$ 0.64  &    40.5   &    59.2 $\pm$ 4.9  &  38.0 $\pm$ 8.9  &  0.19 $\pm$ 0.02  &  0.46 $\pm$ 0.11  \\
DK Tau  &  23.13 $\pm$ 2.96  &  378.29  &  5.64  &  1.33 $\pm$ 1.00  &  63.8    &    69.0 $\pm$ 7.4  &  35.7 $\pm$ 2.2  &  0.03 $\pm$ 0.00  &  0.10 $\pm$ 0.01  \\
DM Tau  &  5.54 $\pm$ 0.71  &  123.69  &  1.68  &  7.82 $\pm$ 2.41  & 13.6   &     44.5 $\pm$ 0.5  &  40.2 $\pm$ 4.5  &  0.29 $\pm$ 0.00  &  0.36 $\pm$ 0.04  \\
DN Tau  &  101.03 $\pm$ 26.50  &  1171.54  &  259.67  &  $\cdots$ $\pm$ $\cdots$  &  2580.2   &     41.9 $\pm$ 7.4  &  49.7 $\pm$ 3.1  &  0.40 $\pm$ 0.07  &  0.29 $\pm$ 0.02  \\
DR Tau  &  21.79 $\pm$ 7.61  &  412.74  &  11.40  &  $\cdots$ $\pm$ $\cdots$  &  2.9~$\times$~10$^{5}$  &    39.4 $\pm$ 2.5  &  26.8 $\pm$ 13.4  &  0.02 $\pm$ 0.00  &  0.04 $\pm$ 0.02  \\
HN Tau  &  15.87 $\pm$ 2.66  &  280.88  &  2.98  &  2.11 $\pm$ 0.73  &   50.3  &    64.1 $\pm$ 4.9  &  53.6 $\pm$ 6.7  &  0.43 $\pm$ 0.03  &  0.62 $\pm$ 0.08  \\
IN Cha  &  0.71 $\pm$ 0.08  &  27.38  &  0.04  &  $\cdots$ $\pm$ $\cdots$  &   0.8  &     61.7 $\pm$ 2.5  &  40.2 $\pm$ 15.6  &  0.13 $\pm$ 0.01  &  0.31 $\pm$ 0.12  \\
IP Tau  &  1.08 $\pm$ 0.16  &  32.38  &  1.07  &  $\cdots$ $\pm$ $\cdots$  &  4.0  &    51.7 $\pm$ 4.9  &  53.6 $\pm$ 35.7  &  0.30 $\pm$ 0.03  &  0.28 $\pm$ 0.18  \\
MY Lup  &  68.89 $\pm$ 5.93  &  941.26  &  19.46  &  $\cdots$ $\pm$ $\cdots$  &   111.8  &    39.5 $\pm$ 2.5  &  40.2 $\pm$ 4.5  &  2.20 $\pm$ 0.14  &  2.12 $\pm$ 0.24  \\
RY Lup  &  10.90 $\pm$ 2.18  &  216.77  &  4.02  &  3.33 $\pm$ 2.01  &   18.7   &     49.3 $\pm$ 2.5  &  42.4 $\pm$ 13.4  &  2.15 $\pm$ 0.11  &  2.90 $\pm$ 0.92  \\
SY Cha  &  12.84 $\pm$ 1.15  &  264.34  &  2.87  &  6.01 $\pm$ 0.81  &   44.5  &   74.0 $\pm$ 2.5  &  55.8 $\pm$ 4.5  &  0.31 $\pm$ 0.01  &  0.55 $\pm$ 0.04  \\
TX Ori  &  11.38 $\pm$ 1.46  &  329.57  &  1.59  &  $\cdots$ $\pm$ $\cdots$  &    263.0   &      56.7 $\pm$ 2.5  &  40.2 $\pm$ 4.5  &  0.90 $\pm$ 0.04  &  1.80 $\pm$ 0.20  \\
UX Tau A  &  8.11 $\pm$ 0.59  &  165.18  &  2.29  &  1.10 $\pm$ 0.58  &    8.8   &    44.4 $\pm$ 2.5  &  44.7 $\pm$ 2.2  &  0.77 $\pm$ 0.04  &  0.76 $\pm$ 0.04  \\
V397 Aur  &  -0.00 $\pm$ 0.05  &  2.22  &  0.45  &  $\cdots$ $\pm$ $\cdots$  &  0.4   &     $\cdots$ $\pm$ $\cdots$  &  $\cdots$ $\pm$ $\cdots$  &  $\cdots$ $\pm$ $\cdots$  &  $\cdots$ $\pm$ $\cdots$  \\
XX Cha  &  8.63 $\pm$ 1.09  &  198.02  &  2.72  &  $\cdots$ $\pm$ $\cdots$  &  27.6   &     51.8 $\pm$ 7.4  &  40.2 $\pm$ 2.2  &  0.29 $\pm$ 0.04  &  0.48 $\pm$ 0.03  \\
V505 Ori  &  49.55 $\pm$ 4.61  &  1063.19  &  2.93  &  $\cdots$ $\pm$ $\cdots$  &  1164.4   &    51.8 $\pm$ 4.9  &  38.0 $\pm$ 13.4  &  0.80 $\pm$ 0.08  &  1.50 $\pm$ 0.53  \\
V510 Ori  &  25.63 $\pm$ 4.81  &  629.94  &  1.62  &  15.50 $\pm$ 5.85  &   174.4   &    41.9 $\pm$ 4.9  &  42.4 $\pm$ 6.7  &  1.15 $\pm$ 0.13  &  1.12 $\pm$ 0.18 
\enddata
\end{deluxetable}
\end{longrotatetable}

\clearpage

\end{document}